\documentclass{jfm} 

\usepackage{graphbox}
\usepackage{pgfplots}
\usepackage{natbib}
\usepackage{changes}
\usepackage[utf8]{inputenc}
\usepackage{amsmath}
\usepackage{bm}
\usepackage{float}
\usepackage{amssymb}
\usepackage{subcaption}
\usepackage{natbib}
\usepackage{mathrsfs, bbm, upgreek}
\usepackage[ruled, vlined]{algorithm2e}
\usepackage[colorlinks=true, allcolors=blue]{hyperref}
\usepackage{xcolor}

\newcommand{\dumarg}[1]		{\widetilde{#1}}

\newcommand{\ie}		{\textit{i.e.}}

\newcommand{\argmin}{\operatornamewithlimits{arg\ min}}
\newcommand{\argmax}{\operatornamewithlimits{arg\ max}}

\mathchardef\mhyphen="2D		

\newcommand{\ddroit}		{{\mathrm{d}}}
\newcommand{\card}[1]		{\mathrm{card}\left[#1\right]}

\newcommand{\transpose}		{\mathsf{T}}

\newcommand{\trace}[1]		{\mathrm{Tr}\left[#1\right]}

\newcommand{\matricize}[1]	{\MakeUppercase{#1}}

\newcommand{\diag}		{\mathrm{diag}}

\newcommand{\be}	{\begin{equation}}
\newcommand{\ee}	{ \end{equation}}
\newcommand{\bi}	{\begin{itemize}}
\newcommand{\ei}	{ \end{itemize}}
\newcommand{\bea}	{\begin{eqnarray}}
\newcommand{\eea}	{ \end{eqnarray}}
\newcommand{\benum}	{\begin{enumerate}}
\newcommand{\eenum}	{ \end{enumerate}}
\newcommand{\bc}	{\begin{center}}
\newcommand{\ec}	{ \end{center}}
\newcommand{\bealign}	{\begin{align}}
\newcommand{\eealign}	{ \end{align}}

\newcommand{\normLM}[2]		{\left\|{#1}\right\|_{#2}}

\newcommand{\R}			{\mathbb{R}}

\newcommand{\bmat}{\begin{bmatrix}}
\newcommand{\emat}{\end{bmatrix}}

\newcommand{\cluster}   {\mathcal{C}}
\newcommand{\loss}          {\mathscr{L}}

\newcommand{\eigenmat}      {\Phi}

\newcommand{\matmode}   {\Phi}

\newcommand{\snap}          {f}
\newcommand{\bsnap}         {\boldsymbol{\snap}}
\newcommand{\snapflu}       {\snap'}
\newcommand{\bsnapflu}      {\bsnap'}

\newcommand{\matsnap}       {\matricize{\snap}}

\newcommand{\Nt}            {{N_s}}
\newcommand{\ntLM}          {\nsnap}
\newcommand{\nsnap}         {{\Nt}}
\newcommand{\isnap}         {{i}}

\newcommand{\imode}         {{n}}

\newcommand{\nmode}         {{N_T}}
\newcommand{\modeLM}        {\phi}
\newcommand{\bmode}         {\boldsymbol{\modeLM}}

\newcommand{\ntopic}        {\nmode}
\newcommand{\topic}         {z}
\newcommand{\btopic}        {\boldsymbol{\topic}}

\newcommand{\corLM}       {C}
\newcommand{\corLMemp}    {\widehat{\corLM}}
\newcommand{\cordual}   { \widetilde{C} }

\newcommand{\eigval}        {\lambda}

\newcommand{\coef}          {a}
\newcommand{\vecoefs}       {\mathbf{\coef}}
\newcommand{\bcoefs}        {\vecoefs}
\newcommand{\coefmat}       {\matricize{\coef}}
\newcommand{\fieldmat}      {\matsnap}
\newcommand{\flufieldmat}  {\fieldmat'}

\newcommand{\tvecA}        {\bcoefs}

\newcommand{\centroid}      {\boldsymbol{c}}
\newcommand{\centro}        {\centroid}

\newcommand{\set}[1]    {\left\{#1\right\}}
\newcommand{\admsetX}       {\mathcal{S}_X}
\newcommand{\admsetY}       {\mathcal{S}_Y}

\newcommand{\Dx}        {\Omega}
\newcommand{\vecx}      {\mathbf{x}}
\newcommand{\ix}        {{l}}
\newcommand{\iunit}        {{j}}
\newcommand{\Nx}        {{N_x}}
\newcommand{\nx}        {\Nx}

\newcommand{\proba}         {p}

\newcommand{\bgrid}         {\mathbf{x}}

\newcommand{\Dirichlet}     {\mathrm{Dir}}
\newcommand{\Dir}           {\Dirichlet}

\newcommand{\vecalpha}      {\boldsymbol{\upalpha}}
\newcommand{\balpha}        {\vecalpha}

\newcommand{\expe}          {\mathbb{E}}

\newcommand{\by}            {\boldsymbol{y}}

\newcommand{\boldone}       {\boldsymbol{1}}
\newcommand{\btheta}        {\boldsymbol{\theta}}

\newcommand{\leftSV}        {\Psi}
\newcommand{\wtdistribscal} {\varphi}
\newcommand{\wtdistrib}     {\boldsymbol{\wtdistribscal}}

\newcommand{\thres}         {\kappa}

\newcommand{\ttheta}{\widetilde{\theta}}
\newcommand{\ta}{\widetilde{a}}
\newcommand{\bbbeta}{{\boldsymbol \beta}}

\newcommand{\nt}{\widetilde{t}}
\newcommand{\tN}{\widetilde{N}}

\usepackage{stackrel}

\begin{document}

\title{Coherent structure identification in turbulent channel flow using Latent Dirichlet Allocation}
\author{Mohamed Frihat, B\'ereng\`ere Podvin, Lionel~Mathelin, Yann~Fraigneau, Fran\c{c}ois Yvon}
\maketitle

\begin{abstract}
Identification of coherent structures is an essential step  to  
describe and model turbulence generation mechanisms in wall-bounded flows.  
To this end, we present a clustering method based on Latent Dirichlet Allocation (LDA),
a generative probabilistic model for collections of discrete 
data.
The method is applied  to  a set of snapshots featuring 
the Reynolds stress {($Q_-$ events)} for a 
turbulent channel flow at a  moderate Reynolds number $R_{\tau}=590$.
Both 2D and 3D analysis show that LDA provides a robust and compact 
flow description in terms of a combination of {\it motifs}, which are latent variables
inferred from the set of snapshots.  
We find that the characteristics of the 
motifs scale with the wall distance, in agreement 
with the wall-attached eddy hypothesis of \citet{kn:townsend61}.
LDA motifs can be used to reconstruct fields with an efficiency 
that can be compared with the POD.
Moreover, the LDA model makes it possible to generate a collection of synthetic fields
that is statistically closer to the original dataset than its POD-generated counterpart.  
These findings highlight the potential of LDA for turbulent flow analysis, compression 
and generation.
\end{abstract}

\section{Introduction}

The introduction of coherent structures \citep{kn:klineandfriends,kn:townsend47}
has represented  a major paradigm shift for   turbulence theory and has had a significant impact 
in various related fields, ranging from geophysical flows to industrial applications.
Coherent structure identification has become a key step towards modelling and controlling wall-bounded turbulent 
flows. However a recurrent stumbling block is the absence of a precise definition of structures, as 
is apparent from several comprehensive reviews  \citep{kn:cantwell81,kn:robinson,kn:jimenez13,
kn:dennis15}.

 Studies originating in the 1960's 
\citep{kn:klineandfriends,kn:kimandfriends} have  established
that most of the turbulence in the near-wall region occurred  in a highly
intermittent manner in both space and time, during what was originally termed ``bursting events''.
 Quadrant analysis of the Reynolds stress in the plane of streamwise and wall-normal fluctuation $(u',v')$
was introduced by \citet{kn:wallabrodkey, kn:willmarthlu72}  to characterize these events. 
Bursting events were  found to be associated
with low-speed streaks being lifted away from the wall, as well with sweeping motions of 
high-speed fluid towards the wall, which respectively correspond to Quadrant II ($u'<0, v'>0)$ and 
Quadrant IV ($u'>0, v'<0)$ events. 
The two quadrants corresponding to $-u'v'>0$ can be termed $Q_-$ events and represent the major contribution
to the Reynolds stress \citep{kn:wallace2016}.
An interpretation of these bursts is that they are the signature of coherent structures or
eddies advected by the mean field. 
Determining the characteristics of these structures has been the 
object of considerable effort, \cite{jimenez18}. 

A central element of wall turbulence theory is the attached eddy model, reviewed 
in detail by \citet{kn:marusicmonty19}.
The model is based on the idea that turbulence arises as a field of randomly distributed eddies, identified as organized flow patterns 
which extend to the wall, in the sense that their characteristics are influenced by the wall. 
Further assumptions require that the entire geometry of the eddies scales with the wall 
distance, with a constant characteristic velocity scale.
The model was extended by \citet{kn:perrychong82}, who introduced
the idea of a hierarchy of discrete scales, with an inverse-scale probability distribution. 
\citet{kn:woodcockmarusic15} showed that this inverse probability
distribution was in fact a  direct consequence of the self-similarity of the eddies. 
Further extensions of the model for the logarithmic layer include a wider variety of structures,
such as wall-detached ones \citep{kn:perry95, hu20}.

Detection of self-similarity in boundary layers has been the focus of 
several experimental studies, such as
\citet{kn:baars17}'s, who used spectral coherence analysis to provide 
evidence of self-similar structures in the 
streamwise velocity fluctuations of pipe flow.
Numerical simulation has proved a powerful tool to explore 
three-dimensional flow fields using a clustering approach.  
Examples include the work of
\citet{kn:delalamo06}, who showed  that
the logarithmic region of turbulent channel was organized in self-similar vortex clusters, and  
\citet{kn:lozanoduran12} developed  a three-dimensional extension of
quadrant analysis  to detect self-similarity in numerical data 
at various Reynolds numbers. 
More recently, wall-attached structures  were identified in the 
streamwise fluctuations  of a turbulent boundary layer
\citep{kn:hwang18} as well as in pipe flow \citep{kn:hwang19}.
The structures were shown to scale with the wall distance while
their population density scales inversely with the distance to the wall.  
\citet{chengcheng2020} detected the signature of  wall-attached eddies 
in the streamwise and
spanwise velocity fluctuations in turbulent channel flow simulations at low  Reynolds numbers.
Evidence of self-similarity has been found as well in the context of resolvent analysis, \cite{sharmamckeon13}. 
It has also emerged  from Proper Orthogonal Decomposition (POD) results, such as channel flow simulations at low Reynolds numbers \citep{kn:jfe10,kn:pof17}, or pipe flow experiments \citep{hellstrom16}. 


The increase of available data, whether through numerical simulation or experiment, has strengthened the need for new identification methods, such as those provided by machine learning
(see \citet{kn:brunton2020} for a review).
The challenge is to extract structural information about the data without pre-existing knowledge,
which defines an {\it unsupervised learning} problem.
Solutions to this problem should be  robust, easy to implement and scalable. 
One example of unsupervised learning method that meets these criteria is 
Proper Orthogonal Decomposition \citep{kn:lumleyPOD}, 
a now classical approach to decompose turbulent fields. 
POD is a statistical technique which provides 
an objective representation of the data as a linear combination of spatial 
eigenfunctions, which can be hierarchized with respect to a given norm.
Although the reconstruction is optimal with respect to this norm
\citep{kn:HLB}, a potential limitation of the decomposition is that
the physical interpretation of the eigenfunctions is not clear. 
In particular, in the case of homogeneous statistics, the eigenfunctions are spatial Fourier modes over 
 the full domain {(see \cite{kn:HLB} for a proof)}, even though 
instantaneous patterns are  strongly localized in space. 
The  connection between POD spatial eigenfunctions 
with observed coherent structures is therefore not necessarily straighforward.
Moreover, the amplitudes of the spatial eigenfunctions are generally strongly inter-dependent, even though 
they are by construction uncorrelated. 
This makes it difficult to give a physical meaning to individual amplitudes, especially in the 
absence of a probabilistic framework in which to interpret them.

In this paper we consider such a framework to explore an alternative
 unsupervised learning approach called 
Latent Dirichlet Allocation (LDA), which can be derived 
from POD \citep{kn:hofmann99}.
LDA is a generative probabilistic model, that is a probabilistic model that 
mimics the characteristics of a collection of data.
  It is based on a soft clustering approach,
which was first developed for text mining applications \citep{kn:blei03}, but has been extended to other 
fields in recent years \citep{kn:aubert13}. 
The goal of LDA \citep{kn:blei03}  is to find short descriptions of the members of a collection that enable efficient processing of large collections while preserving the essential statistical relationships that are useful for basic tasks such as classification, novelty detection, summarization, and similarity and relevance. LDA is a three-level hierarchical Bayesian model, in which each member of a collection is modeled as a finite mixture over an underlying set of topics or motifs. 

In the field of natural language processing, the dataset to which LDA is applied 
consists of a set of documents, each of which
is considered as a ``bag-of-words'', that is an unordered set of words taken from a finite vocabulary.
A particular word may appear several times in the document, or not appear at all.
The number of occurrences  of each vocabulary word  in a document can be seen as an entry 
of a sparse matrix where the lines correspond to the vocabulary  words 
and the columns to the documents. 
Based on this typically sparse word count matrix, the classification method returns a 
set of $\ntopic$   {\it topics}, where the topics  are latent variables 
inferred { from the word counts in the documents}
and the number of topics $\ntopic$ is a user-defined parameter.

Unlike ``hard'' clustering, such as the K-means approach \citep{kn:macqueen1967}, where 
each document is assigned to a specific topic, LDA represents each document 
as { a mixture of topics, where the coefficients of the mixture represent the probability
of the topic in the document.}
An interesting application of the LDA method was carried out 
for { a dataset containing} images by \citet{kn:griffiths04}.
The dataset considered 
was  a collection of gray-scale images where each image consists of an array of pixels, each of which is associated with a gray level. In this framework, 
each image is the equivalent of a document, 
each pixel represents an individual vocabulary word, and 
the gray-level intensity measured at each 
pixel is taken as the analog of the word count matrix entry (the lines of the 
matrix now represent the pixels, while the columns represent the snapshots).
The sum of the intensities over the pixels, which will be called throughout
the paper the {\it total intensity},  is the analog of the total number of
words observed in the document. 
Given a set of original patterns constituting the topics or {\it motifs}, a 
collection of synthetic images was generated from random mixtures of the 
patterns.
It was shown that LDA was able to recover 
the underlying patterns from the observations of
the generated images. 

Following \cite{kn:griffiths04}, the idea of the paper 
is to look for evidence of coherent structure in turbulent flow snapshots 
by identifying LDA topics or {\it motifs}.
The relevant gray-level intensity is based on the value of $Q_-$
(unlike in \citet{kn:griffiths04}'s work, it corresponds to a physical field.)
We thus propose the following analogy: each  
 scalar field observed in a collection of snapshots results 
from a mixture of $\ntopic$ 
spatial {\it topics} 
{that will be referred to as {\it motifs} in the remainder of the paper}.
This can be compared with the standard view  
that each realization of a turbulent flow is constituted of 
a random superposition of discrete eddies, characterized by a hierarchy 
of scales.

The paper is organized as follows. 
We show in Section~\ref{Sec_decomp} how the POD method of snapshots, which
is equivalent to Latent Semantic Allocation (LSA), can be  generalized 
to a probabilistic  framework (Probabilistic Latent Semantic Allocation
or PLSA) which is then further extended into Latent Dirichlet Allocation
(LDA) in Section~\ref{Sec_LDA}.   
Application to the extraction of motifs for a turbulent channel flow is introduced in Section~\ref{Sec_channel} and results are discussed in Section~\ref{Sec_results}. The potential of the approach for flow reconstruction and flow generation is considered in Section~\ref{Sec_reconstruction} before Section~\ref{Sec_Conclusion} closes the paper.

\section{A probabilistic extension of Proper Orthogonal Decomposition } \label{Sec_decomp}
To suitably introduce and contextualize the Latent Dirichlet Allocation, several established approaches to represent data are first briefly discussed.

%
%
\subsection{Proper Orthogonal Decomposition}

\subsubsection{General formulation}

The Proper Orthogonal Decomposition (POD) is arguably the most popular tool for representation and analysis of turbulent flow fields. It relies on a method rediscovered and revisited several times in different scientific domains and comes by several names (Principal Component Analysis, Empirical Mode Decomposition, Karhunen-Lo\`eve decomposition, Latent Semantic Allocation (LSA) \ldots) although they are not all strictly equivalent. It was introduced for turbulent flows and adapted by \citet{kn:lumleyPOD}.

The POD method allows to derive an orthogonal basis for the (sub)space of the fluctuations of a multi-dimensional quantity $\bsnap$ of finite variance.
One can show that a basis for the space of fluctuations, defined as $\bsnapflu\left(t\right) := \bsnap\left(t\right) - \left<\bsnap\right>$, with $\left<\cdot\right>$ the statistical mean, is given by the set of elements $\set{\bmode_\imode}_\imode$, eigenvectors of the following eigenvalue problem \citep{kn:HLB}:  
\be
\corLM \bmode_\imode = \eigval_\imode \bmode_\imode, \label{Eq_eigpb_POD}
\ee
with $\eigval_\imode$ the eigenvalue and $\corLM \in \R^{\nx \times \nx}$ the empirical 2-point covariance matrix:
%
\be
\corLM = \frac{1}{\ntLM} \sum_{\isnap=1}^{\nsnap}{\bsnapflu\left(t_\isnap\right) \bsnapflu\left(t_\isnap\right)}, \label{Eq_empicov}
\ee
with $\set{t_\isnap}_\isnap$ the time instants for which the field $\bsnap$ is available. 
Some conditions on the temporal sampling scheme apply for the empirical covariance $\corLMemp$ to be an accurate approximation of $\corLM$ \citep{kn:HLB}. POD modes are identified as the eigenvectors $\bmode_\imode$.


\subsubsection{Method of snapshots}

The above method is a quite natural implementation of the underlying Hilbert-Schmidt decomposition theory. However, the algorithmic complexity associated with the eigenvalue problem \eqref{Eq_eigpb_POD} scales as $\mathcal{O}\left(\ntLM \, \nx^2\right)$, where the number of field instances $\ntLM$ was assumed to be lower than the size $\nx$ of the discrete field, $\ntLM \le \nx$. For large field vectors (large $\nx$), the computational and memory cost is hence high. For this widely encountered situation, a possible workaround was suggested in \cite{kn:siro87} and consists in solving the following eigenvalue problem:
\be
\cordual \, \tvecA_\imode = \eigval_\imode \tvecA_\imode, \qquad \tvecA_\imode \in \R^\ntLM, \label{Eq_eigpb_POD_snapshot}
\ee
with
\be
\cordual_{\isnap, {\isnap'}} \propto \left<\bsnapflu\left(t_\isnap\right), \bsnapflu\left(t_{\isnap'}\right)\right>_\Dx, \qquad \forall \: \isnap, {\isnap'} \in \left[1, \ntLM\right] \subset \mathbb{N},
\label{defcordual}
\ee
and $\left<\cdot, \cdot\right>_\Dx$ the Euclidean inner product. 
Since the correlation matrix $\cordual$ is Hermitian, its eigenvalues are real and non-negative, $\eigval_\imode \ge 0$, $\forall \, \imode$, and its eigenvectors $\set{\tvecA_\imode}_\imode$ are orthogonal and can be made orthonormal in an Euclidean sense, $\tvecA_\imode^\transpose \, \tvecA_{\imode'} \propto \delta_{\imode, \imode'}$, with $\delta$ the Kronecker delta. 
The spatial POD modes
are finally retrieved via projection as follows:
\be
\bmode_\imode = \eigval_\imode^{-1\slash 2} \, \flufieldmat \, \tvecA_\imode, \qquad \forall \, \imode.
\ee
where the $\isnap$-th column of the matrix $\flufieldmat$ is the 
snapshot $\bsnapflu_\isnap$. 

The algorithmic complexity is now $\mathcal{O}\left(\ntLM^3\right)$ and scales much better than the standard POD approach ($\mathcal{O}\left(\ntLM \, \nx^2\right)$) in the usual situation where $\ntLM \ll \nx$.
In this work, we rely on this so-called method of snapshots to implement POD.

Formally the decomposition of the snapshot matrix $\flufieldmat$ is equivalent to
a singular value decomposition SVD
\be
\flufieldmat = \matmode \Sigma \coefmat^\transpose,
\label{lsa}
\ee  
where $\matmode$ is the matrix constituted by the $\imode$ columns $\bmode_\imode$,

$\coefmat$ is the matrix containing the $\imode$ columns $\tvecA_\imode$  
and $\Sigma$ is a diagonal matrix whose entries are  $\eigval_\imode^{-1\slash 2}$.
The snapshot matrix can thus be decomposed into a  snapshot-mode  matrix $A$
and into a cell-mode  matrix $\matmode$. 
The spatial modes or {\it structures} can be seen as latent variables 
allowing optimal reconstruction of the data in the $L_2$ norm or an equivalent.
The decomposition can be truncated to retain only the $\nmode$ largest values 
corresponding to the $\nmode$ first columns of each matrix.

\subsection{Probabilistic Latent Semantic Analysis} \label{Sec_PLSA}
In all that follows we
will consider a collection of $\nsnap$ scalar fields $\{\bsnap_\isnap \}_{\isnap=1, \cdots, \nsnap}$. Each field is  of dimension $\nx$
and consists of either positive or zero integer values on each grid cell.
For each snapshot $\isnap$,
the value of $\bsnap_\isnap$ on grid cell $\ix$  indicates that
the grid cell $\isnap$ has been detected or activated $\snap_{\ix,\isnap}$ times.
Probabilistic Latent Semantic Analysis (PLSA) tackles the problem of
finding latent variables using a probabilistic method instead of SVD.
This representation assumes that each snapshot $\bsnap_\isnap$ consists
of a mixture of structures $\btopic_\imode$.

PLSA adds a probabilistic flavor as follows:
\begin{itemize}
    \item given a snapshot $\bsnap_\isnap$, the structure $\btopic_\imode$ is present in that snapshot with probability $\proba(\btopic_\imode|\bsnap_\isnap)$,
    \item given a structure $\btopic_\imode$, the grid cell $\bgrid_\ix$ is activated
with probability $\proba(\bgrid_\ix|\btopic)$.
\end{itemize}

Formally, the joint probability of seeing a given snapshot $\bsnap_\isnap$ and
activating a grid cell $\bgrid_\ix$ is:
\begin{equation}
\proba(\bsnap_\isnap,\bgrid_\ix)= \proba(\bsnap_\isnap)\sum_{\imode} \proba(\btopic_\imode|\bsnap_\isnap) \proba(\bgrid_\ix|\btopic_\imode).
\label{Eq_PLSAone}
\end{equation}

$\proba(\bsnap_\isnap)$, $\proba(\btopic_\imode|\bsnap_\isnap)$, and $\proba(\bgrid_\ix|\bsnap_\isnap)$ are the parameters of the model: $\proba(\bsnap_\isnap)$ is the probability to obtain such a snapshot $\bsnap_\isnap$ and is constant in our case, $\proba(\bsnap_\isnap)=1/\nsnap$. $\proba(\btopic_\imode|\bsnap_\isnap)$ and $\proba(\bgrid_\ix|\btopic_\imode)$ can be infered using the Expectation-Maximization (EM) algorithm of \citet{dempster77}.

Using Bayes' rule, $\proba(\bsnap_\isnap, \bgrid_\ix)$ can be equivalently written as: 
\begin{equation}
\proba(\bsnap_\isnap,\bgrid_\ix)= \sum_{\imode} \proba(\btopic_\imode) 
\proba(\bgrid_\ix|\btopic_\imode) \proba(\bsnap_\isnap|\btopic_\imode). 
\label{Eq_PLSAtwo}
\end{equation}

%
This alternative formulation shows a direct link between PLSA model and 
POD model (as mentioned above, POD is called Latent Semantic Allocation or LSA in text mining). 
If we compare equations~\eqref{lsa} and \eqref{Eq_PLSAtwo}, we see that 
the structure probability $\proba(\btopic_\imode)$ corresponds to the diagonal matrix $\varLambda_\imode$, the probability of the snapshot $\bsnap_\isnap$  given the structure $\btopic_\imode$ corresponds to the snapshot-mode matrix entry $A_{\isnap, \imode}$, and the probability to activate the cell $\bgrid_\ix$ given the structure $\btopic_\imode$ corresponds to the  matrix entry $\matmode_{\ix, \imode}$.

\section{Latent Dirichlet Allocation} \label{Sec_LDA}


Latent Dirichlet Allocation (LDA) extends PLSA to address its limitations. 
Its specificity is:
\begin{itemize}
\item  the introduction of a probabilistic model for the collection of 
snapshots: each snapshot is now characterized by a distribution over the structures 
 which will be now called {\it motifs}.

\item the use of Dirichlet distributions to  model both motif-cell and  snapshot-motif 
distributions.
\end{itemize}

The Dirichlet distribution is a multivariate probability distribution over the space of 
multinomial distributions. 
It is parametrized by a vector of positive-valued parameters $\balpha=\left(\alpha_1,\ldots,\alpha_N\right)$:
\[ \proba\left(x_1, \ldots, x_N ; \alpha_1, \ldots, \alpha_N\right) =\frac{1}{B(\balpha)}
\prod_{\imode=1}^{N} x_\imode^{\alpha_\imode-1}, \]
where $B$ is a normalizing constant, which can be expressed in terms of the Gamma
function $\Gamma$:
\[ B(\balpha)= \frac{\prod_{\imode=1}^{N} \Gamma(\alpha_\imode)}{\Gamma(\sum_{\imode=1}^{N} \alpha_\imode)}. \]
The support of the Dirichlet distribution is the set of $N$-dimensional discrete distributions,
which constitutes the $N-1$ simplex. 
Introduction of the Dirichlet distribution  allows us to specify the prior belief about the snapshots. The 
Bayesian learning problem is now to estimate 
$\proba(\btopic_\imode, \bsnap_\isnap)$ and $\proba(\vecx_\ix, \btopic_\imode)$ 
 from $\matsnap$ given our prior belief $\balpha$, and it can be shown that 
Dirichlet distributions offer a tractable, well-posed solution
to this problem \citep{kn:blei03}.

LDA is therefore based on the following representation: 
 \begin{enumerate}
    \item Each motif $\btopic_\imode$ is associated with a multinomial distribution $\wtdistrib_\imode$ over the grid cells ($\proba\left(\vecx_\ix|\btopic_\imode\right)= \wtdistribscal_{\ix, \imode} $). 
This distribution is modeled with a Dirichlet prior parametrized with a $\nx$-dimensional vector {\bf $ \bbbeta$}. 
 The components $\beta_\ix$ of $\bbbeta$ control the sparsity of the distribution: values of $\beta_\ix$ larger than 1 correspond to evenly dense distributions, while values lower than 1 correspond to sparse distributions.
 In all that follows, we will assume a  non-informative prior, 
  meaning that $\bbbeta = \beta \boldone_\nx$.

    \item Each snapshot $\bsnap_\isnap$, is associated with a distribution of motifs 
$\btheta_\isnap$ such that $\theta_{\imode,\isnap} = \proba(\btopic_\imode|\bsnap_i)$. 
The probabilities of each motif add up to $1$ in each snapshot. This distribution is modelled with a $\nmode$-dimensional Dirichlet distribution of parameter $\balpha$. The magnitude of $\balpha$ characterizes the sparsity of the distribution (low values of $\alpha_\imode$ correspond to snapshots with relatively few motifs). The same assumption of a non-informative prior leads us to assume $\balpha=\alpha \boldone_\nmode$.  \end{enumerate}



The generative process performed by LDA with $\ntopic$ motifs is the following: 
 \begin{enumerate}
    \item For each motif $\btopic_\imode$, a cell-motif distribution $\wtdistrib_\imode$ is drawn from the Dirichlet distribution of parameter $\beta$. 
    \item For each snapshot $\bsnap_\isnap$:
    \begin{itemize}
        \item a snapshot-motif distribution $\btheta_{\isnap}$ is drawn.
        \item each intensity unit $1 \le \iunit \le N_\isnap$ 
where $N_\isnap$ is the total intensity with 
$N_\isnap= \sum_{\ix} \snap_{\ix, \isnap}$ is then distributed among the different cells as follows:
        \begin{itemize}
            \item a motif $\btopic_{\imode}$ is first selected from $\btheta_\isnap$ (motif $\btopic_\imode$ occurs with probability $\theta_{\imode, \isnap}$ in the snapshot),
            \item for this motif, a cell $\ix$ is chosen among the
cells using $\wtdistribscal_{\ix, \imode}$ and the intensity 
associated with cell $\ix$ is incremented by 1. 
        \end{itemize}
    \end{itemize}
\end{enumerate}

The generative process can be summarized as follows:

\begin{algorithm}[H]
\SetAlgoLined
\For {each of the $\ntopic$  {motifs} $\imode$} 
 { sample $\wtdistrib_\imode \sim \Dir(\beta)$ } \ 
\For {each of the $\nsnap$  {snapshots} $\isnap$}  {  
   sample $\btheta_\isnap \sim \Dir(\alpha)$ \\  
   \For {each of the $N_\isnap$ {intensity units}  }  { 
     1.  sample a  motif $\btopic_{\imode}$ from $\theta_{\imode, \isnap}$ \\
     2.  for this motif sample a cell  $\ix$ from $\wtdistribscal_{\ix, \imode}$ \ 
    }} 
\caption{LDA Generative Model.}
\end{algorithm}

The snapshot-motif distribution $\btheta_{\isnap}$ and the cell-motif distribution $\wtdistrib_\imode$ are determined from the observed $\bsnap_\isnap$. They are respectively $\nmode$- and $\nx$-dimensional categorical distributions.
Finding the distributions $\btheta_{\isnap}$ and $\wtdistrib_\imode$ that are most compatible with the observations is an inference problem that can be solved by either a variational formulation \citep{kn:blei03} 
or a Gibbs sampler \citep{kn:griffiths02}. 
In the variational approach, the objective function to minimize is
the Kullback-Leibler divergence. 
The solution {\it a priori} depends on the number of motifs and on the values of the Dirichlet parameters $\alpha$ and $\beta$.

We conclude this section with two remarks. 
\begin{enumerate}
\item LDA can generalize to new snapshots more easily than PLSA, due to the snapshot-motif distribution formalism.  In PLSA, the snapshot probability is a fixed point in the dataset,
which cannot be estimated directly if it is missing. 
In LDA, the dataset serves as training data for the Dirichlet distribution of snapshot-motif distributions. If a snapshot is missing, it can easily 
be sampled from the Dirichlet distribution instead. 
\item An alternative viewpoint can also be adopted in interpreting the LDA in the form of a regularized matrix factorization method. This is further discussed in Appendix~\ref{Sec_MF}.
\end{enumerate}

\section{Application of LDA to turbulent flows} \label{Sec_channel}

\subsection{Numerical configuration }

\begin{figure}
\centerline{
\includegraphics[height=5cm]{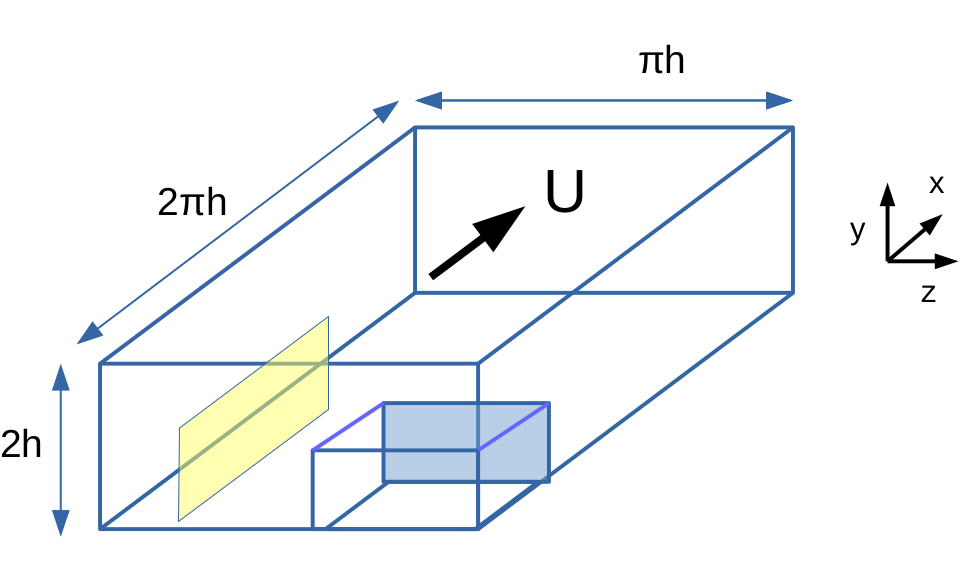}}
\caption{Numerical domain $D$. The shaded surfaces correspond to the two 
types of planes used in the analysis. The volume considered for 3D analysis is indicated
 in bold lines.}
\label{config}
\end{figure}

The idea of this paper is to apply this methodology to snapshots of turbulent flows
in order to determine  latent motifs from observations of
$Q_-$ events.
We will consider the configuration of turbulent channel 
flow at a moderate Reynolds number of $R_{\tau}= u_{\tau} h/\nu= 590$ \citep{kn:moser99,kn:muralidhar19},
where $R_{\tau}$ is the Reynolds number based on the fluid viscosity 
$\nu$, channel half-height $h$ and friction velocity $u_{\tau}$. 
Wall units based on the friction velocity and fluid viscosity  will be denoted with a subscript $_+$.
The streamwise, wall-normal and spanwise directions will be referred to as $x,y$ and $z$ respectively.
The horizontal dimensions of the numerical domain are $(\pi, \pi/2)h$.  
Periodic boundary conditions are used in the horizontal directions. 
The resolution of $(256)^3$ points is based on a regular spacing in the horizontal directions and a hyperbolic
tangent stretching function for the vertical direction. 
The configuration is shown in Figure~\ref{config}.  
More details about the numerical simulation can be found in \citet{kn:muralidhar19}.

\subsection{LDA inputs}

In this section, we introduce the different parameters of the study.
The python library \texttt{scikit-learn} \citep{scikit-learn} was used to 
implement LDA. 
The sensitivity of the results to these parameters will be examined in a subsequent section.

We first focus on 2-D vertical subsections of the domain, then  present  3-D results.
The vertical extent of the domain of investigation was the half-channel height.
Since this is an exploration into a new technique, a 
limited range of scales was considered in the horizontal dimensions:
the spanwise dimension of the domain was limited to 450 wall units. 
The streamwise extent of the domain was in the range  of  450-900 wall units.
The number of realizations considered for 2-D analysis was $\nsnap=800$, with a time separation of 
60 wall time units. The number of snapshots was increased to 2400 for 3-D analysis. 

The scalar field $\bsnap$ of interest corresponds to $Q_-$ events. It 
is defined as the positive part of the product $-u'v'$ , where fluctuations
are defined with respect to an average taken over all snapshots and horizontal planes.
The LDA procedure requires that the input field consists of integer values: 
it was therefore rescaled and digitized and the scalar field $\snap$ was defined as:
\[ \snap = [A \tau_{-}], \] where $\tau_-= \mathrm{max}\left( -u'v',0 \right)$ and $[\cdot]$ represents the integer part. 
The rescaling factor $A$ was chosen in order to yield a sufficiently large, yet still tractable, total intensity.
In practice we used $A=40$, which led to a  total intensity 
$\sum_\isnap \sum_{\ix} \snap_{\ix,\isnap}$ of about $10^8$ for plane sections.
The effect of the rescaling factor will be examined in a subsequent section.  

LDA is characterized by a user-defined  number of motifs $\ntopic$, a parameter $\alpha$ which characterizes
the sparsity of prior Dirichlet snapshot-motif distribution,  and a parameter $\beta$ 
which characterizes
the sparsity of the prior Dirichlet motif-cell distribution.
Results were obtained assuming uniform priors  for $\alpha$ and $\beta$ with a default value
of $1/\ntopic$. The sensitivity of the results to the priors will be evaluated
in Section~\ref{Sec_sensitivity}.

\subsection{LDA outputs}
For a collection of $\nsnap$ snapshots and a user-defined number of motifs $\ntopic$, LDA 
returns $\ntopic$ motif-cell distributions $\wtdistrib_\imode$ and $\nsnap$ snapshot-motif 
distributions $\btheta_\isnap$.
Each {\it motif} is defined by a probability distribution $\wtdistrib_\imode$ associated with each grid cell.
It is therefore analogous to a structure or a portion of structure since 
it contains spatial information - note however that its definition is different from 
standard approaches. 
The motif-snapshot distribution $\btheta_\isnap$ characterizes the prevalence of a given motif in the 
snapshot.

As will be made clear below, the motifs most often consist of single connected regions,
although occasionally a couple of distinct regions were identified.
In most cases, the motifs can thus be characterized by a characteristic location $\vecx^c$ and a
characteristic dimension in each direction $L_j$, $j \in \{x,y,z\}$. 
   
To determine these characteristics, we first define for each motif a mask associated with 
a domain $D$. 
The origin of the domain was defined as the position 
$\vecx_\mathrm{m})$ corresponding to its maximum 
probability $\proba_\mathrm{m} = \wtdistrib_\imode (\vecx_\mathrm{m})$.  The dimensions of the domain in each direction (for instance $L_x$) were 
defined as the segment extending from the domain origin over which the probability remained 
larger than $1\%$ of its maximum value $\proba_\mathrm{m}$.
The position and characteristic dimension of a motif for instance 
in the $x$-direction are then defined as:
\begin{eqnarray}
 x^c & = & \frac {\int_{D} x    \wtdistrib_\imode \ddroit D }{\int_{D}   \wtdistrib_\imode \ddroit D  }, \\ 
 L_x^2 & = &  2 \frac {\int_{D} (x - x^c)^2  \wtdistrib_\imode \ddroit D 
}{\int_{D}  \wtdistrib_\imode  \ddroit D }.    \label{defLx}
\end{eqnarray}
Analogous definitions can be given for $y^c$ and $z^c$.









\section{Results} \label{Sec_results}

\subsection{Vertical planes}

In order to investigate in detail the vertical organization of the flow, 
 LDA was first applied to vertical sections of the flow. 
Both cross-flow $(y,z)$ and longitudinal $(x,y)$ sections were considered.
Due to the horizontal homogeneity of the flow,
we do not expect significant changes in  
the cell-motif and the motif-document distributions 
when the sections are translated in the horizontal direction.

\subsubsection{Cross-flow planes}

\begin{figure}
\includegraphics[height=3.5cm]{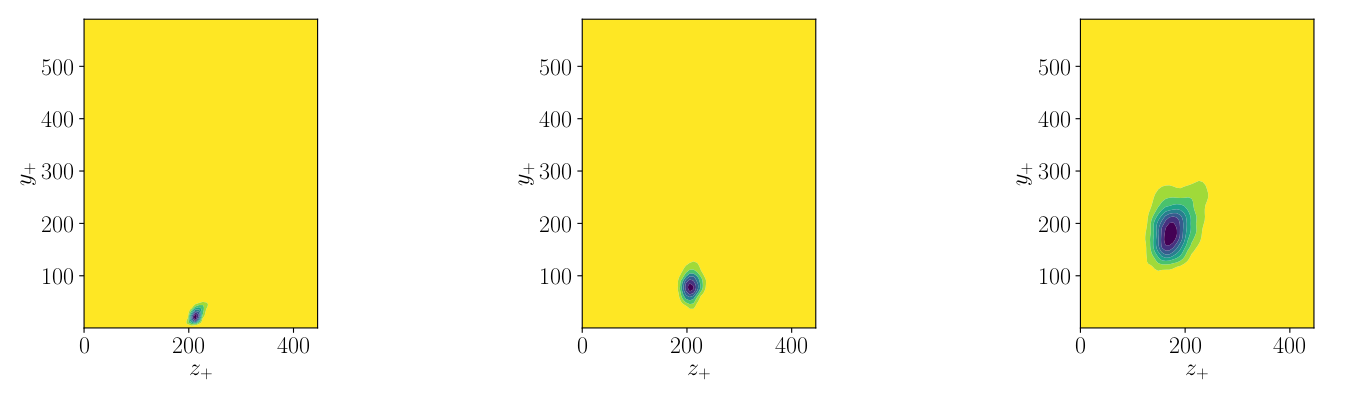} \\
\includegraphics[height=1.1cm]{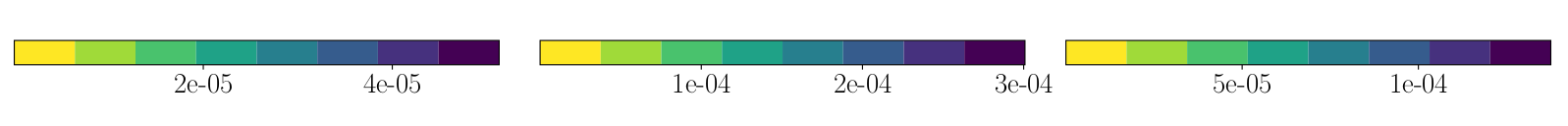} \\
\includegraphics[trim=0.2cm 0 0 0, clip, height=3.5cm]{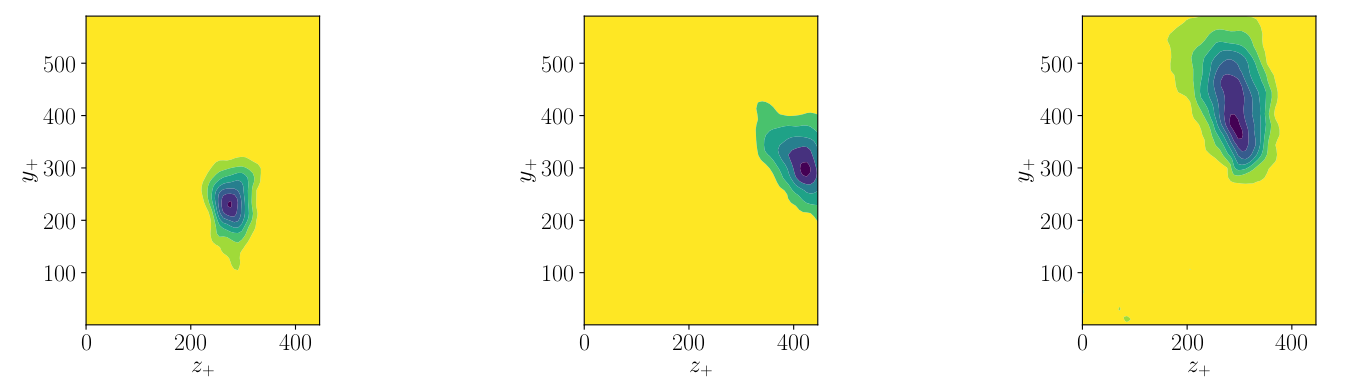} \\
\includegraphics[height=1.1cm]{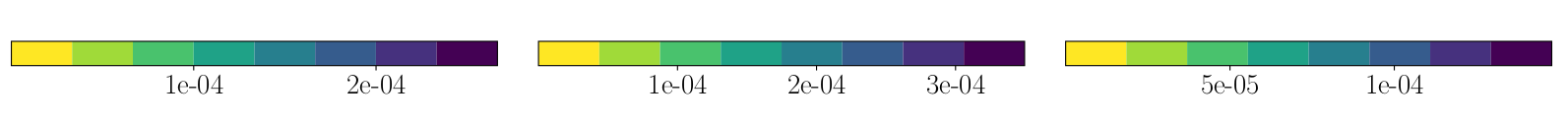} \\
\caption{Selected motifs in a cross-flow plane for a number of motifs  $\ntopic=96.$ }
\label{verticaltopic}
\end{figure}

\begin{figure}
\begin{tabular}{ll}
\includegraphics[height=5cm]{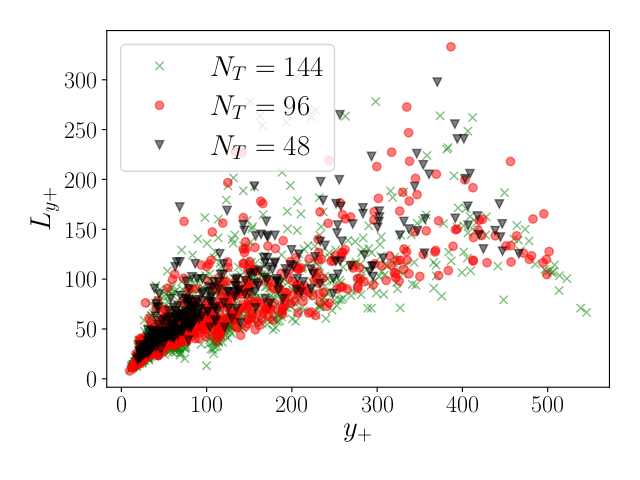} &
\includegraphics[height=5cm]{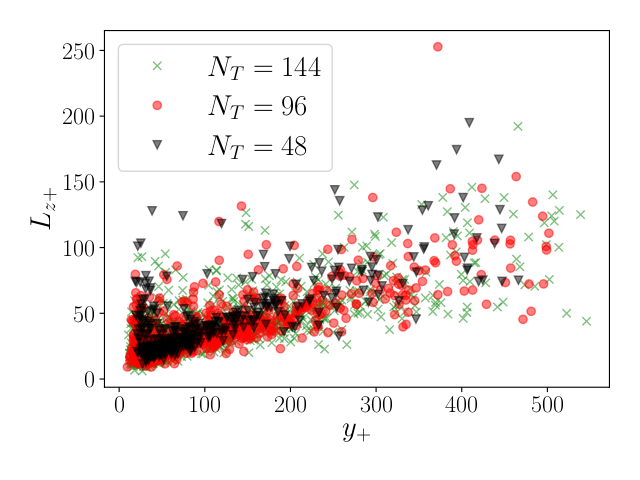} \\ 
\includegraphics[height=5cm]{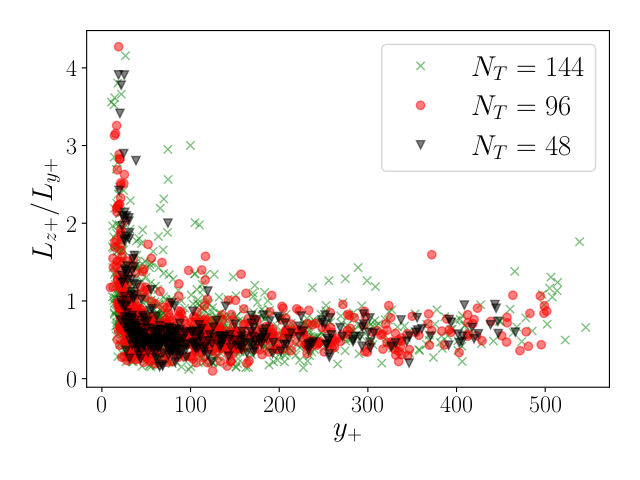} & 
\\ \end{tabular}
\caption{ Cross-plane motif characteristic sizes; 
Left: Vertical dimension $L_y$; Right : Spanwise dimension $L_z$; Bottom: Aspect ratio $L_y/L_z$. 
Each dot corresponds to a motif.}
\label{LyLzvert}
\end{figure}

The dimensions of the cross-sections were $d_{z+}=450$ and $d_{y+}=590$.
Figure~\ref{verticaltopic} shows selected motifs for  a total 
number of motifs $\ntopic=96$ on a vertical plane at $x=0$.  
The motifs consist of isolated regions, the dimensions of which
 increase  with the wall distance. 
This is confirmed by Figure~\ref{LyLzvert}, which represents  
characteristic sizes of LDA motifs  of 
a succession of four vertical planes separated by 
a distance of $100$ wall units (+). 
We point out that observing motifs which are detached from the wall 
does not infirm the presence of wall-attached structures,
as they would be consistent with a cross-section of a
wall-attached structure elongated in the streamwise direction.  
Results for several motif numbers (three different 
motif numbers $\ntopic=48, 96, 144$ are shown in Figure~\ref{LyLzvert}), it was found  
that both spanwise and vertical dimensions increase linearly with 
the wall distance in the region $y_+> 100$.
Again, this is in agreement with \citet{kn:townsend61}'s hypothesis of a hierarchy of structures 
of increasing dimensions, which was also confirmed numerically by \citet{kn:flores10a}.  

The aspect ratio $L_z/L_y$  is constant with the wall distance above $y_+>100$, with a typical value of about 1. 
We note that \citet{kn:lozanoduran12} found with a different definition that $Q_-$ events were characterized by nearly 
equal spanwise and vertical sizes  $\Delta z \sim \Delta y$,  
while \citet{kn:delalamo06} found a scaling of $\Delta z \sim 1.5 \Delta y$
for vortex clusters.

\begin{figure}
\includegraphics[height=5cm]{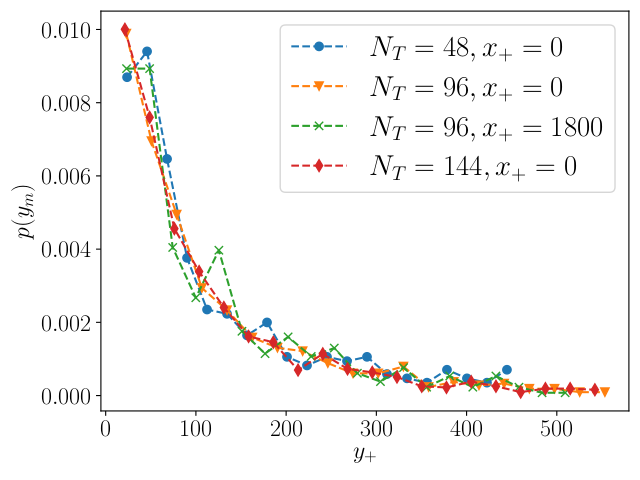} 
\includegraphics[height=5cm]{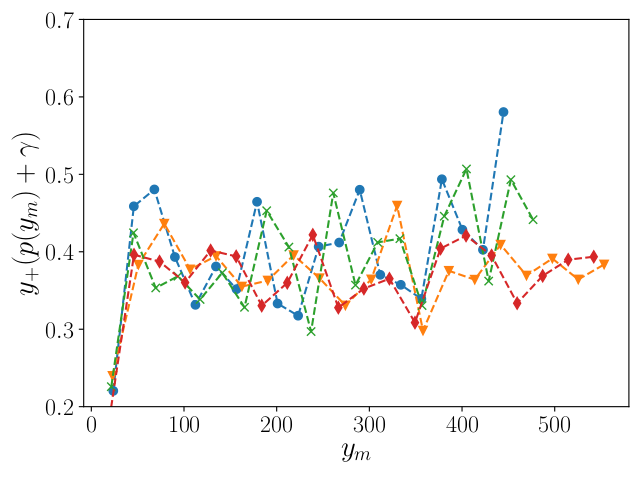}
\caption{Left: Distribution of the motif maximum location $y^c$; Right: Compensated plot of the distribution for different 
sets of motifs and different subdomains. 
The legend is the same for the two figures.}
\label{histzvert}
\end{figure}

Figure~\ref{histzvert} (left) shows the distribution of the vertical location $\proba(y_\mathrm{m})$ of the motif 
maximum probability.
Comparison of two different plane locations $x$ 
confirms that results do not depend on the location of the plane,
which reflects the statistical homogeneity of the flow 
in the horizontal direction.
The probability decreases as the inverse of the wall distance on 
all planes.
This is in agreement with Townsend's self-similarity hypothesis
that the number of structures decreases with the wall distance in $1/y$ 
\citep{kn:townsend61,kn:woodcockmarusic15}. 
Figure~\ref{histzvert} (right)
shows that a good fit is $\proba(y) \simeq \frac{c}{y} - \gamma$, with $\gamma=0.0006$
and $c = 0.4$.

\subsubsection{Longitudinal planes}

\begin{figure}
\begin{tabular}{ll}
\includegraphics[trim={0.8cm 3cm 0 3cm}, clip, height=3.5cm]{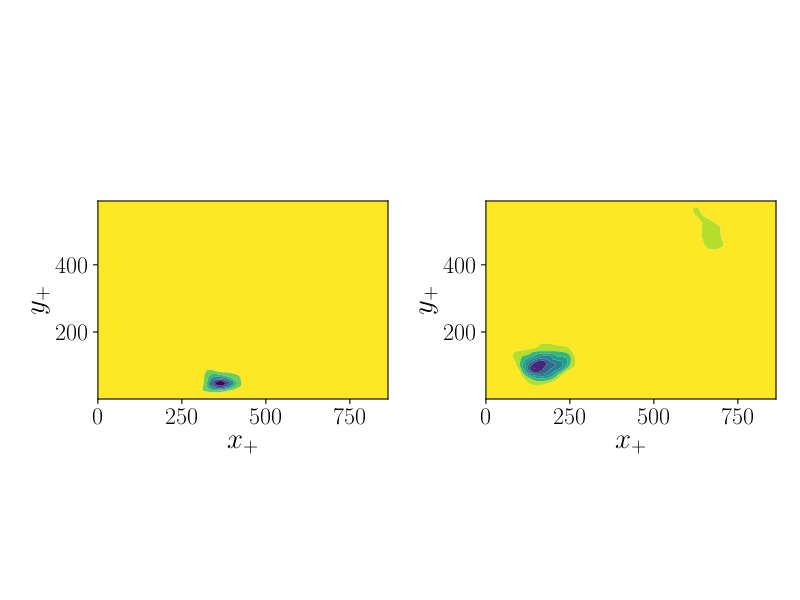} &
\includegraphics[trim={0.8cm 3cm 0 3cm}, clip, height=3.5cm]{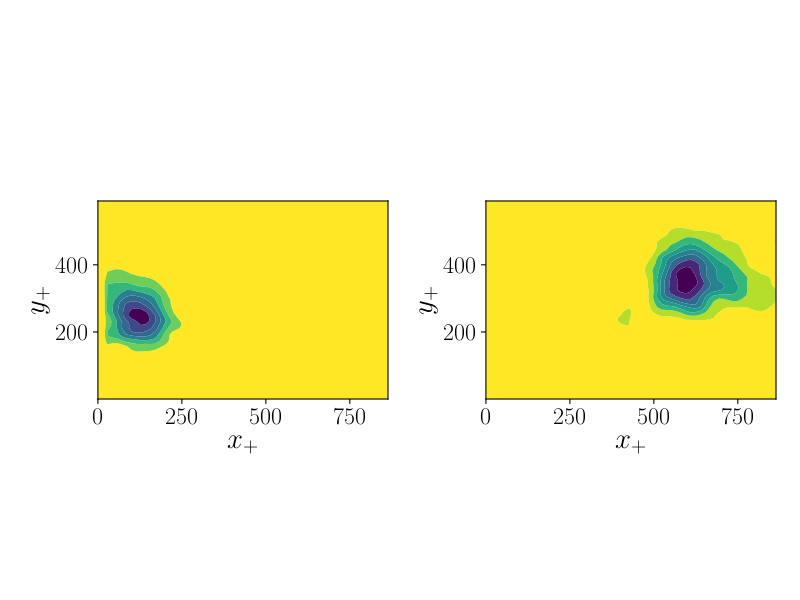} \\ 
\includegraphics[trim={0 0cm 0 12cm}, clip, height=1.2cm]{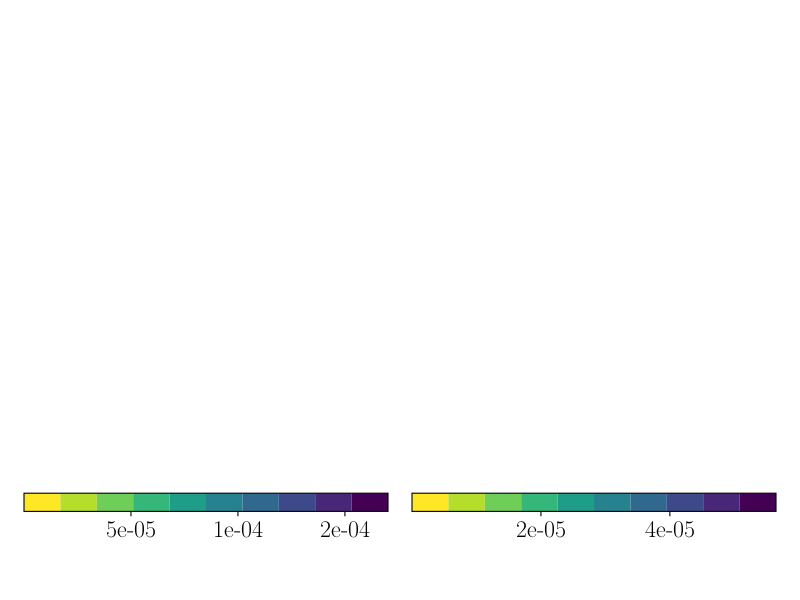} &
\includegraphics[trim={0 0cm 0 12cm}, clip, height=1.2cm]{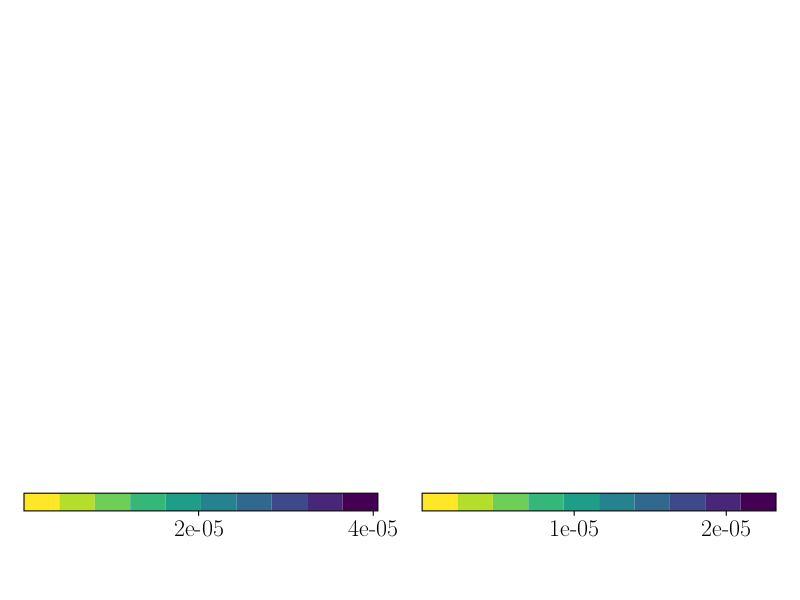} \\ 
\end{tabular}
\caption{Selected motifs for a longitudinal plane with $\ntopic=96$ motifs.}
\label{longtopic}
\end{figure}

\begin{figure}
\begin{tabular}{ll}
\includegraphics[height=5cm]{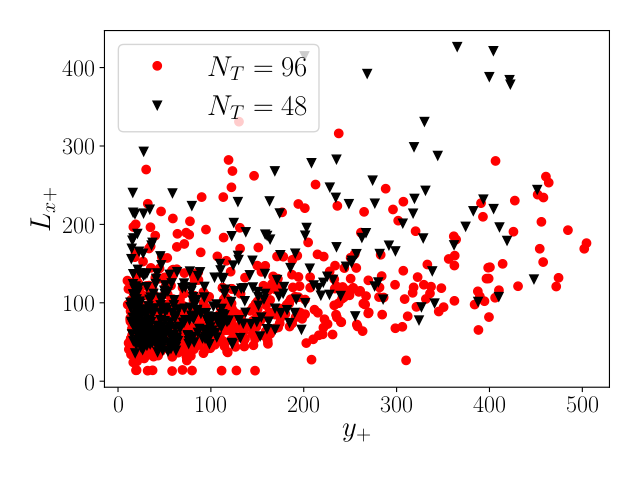} &
\includegraphics[height=5cm]{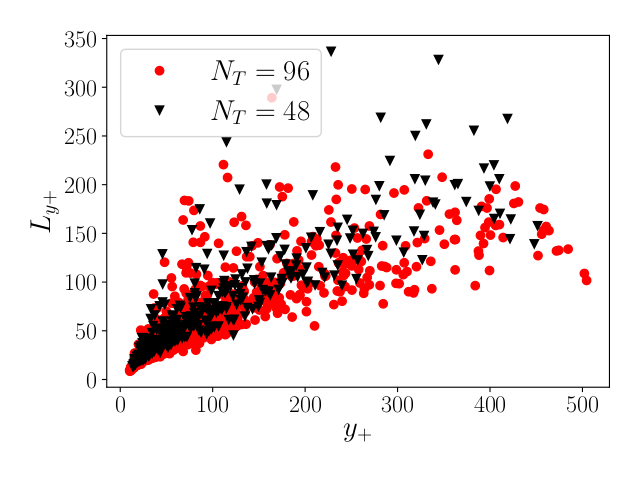} \\ 
\includegraphics[height=5cm]{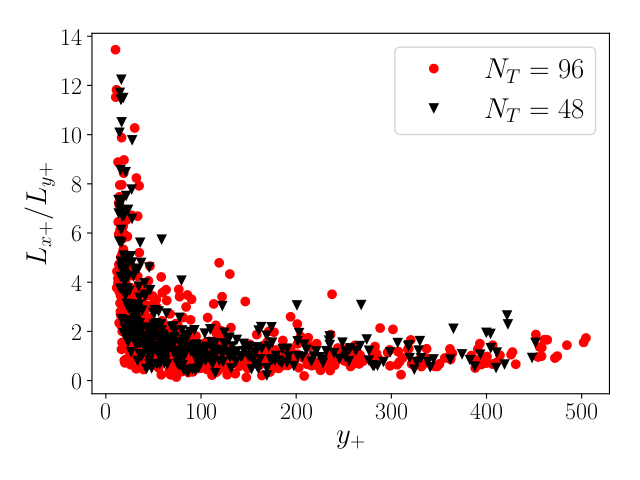} & \\
\end{tabular}
\caption{ Longitudinal motif characteristic dimensions; 
Left: Streamwise dimension $L_x$; Right: Vertical dimension $L_y$; Bottom :  Aspect Ratio $L_x/L_y$. 
Each dot corresponds to a motif.}
\label{LxLylong}
\end{figure}

We now examine results for the longitudinal sections $(x,y)$.
The streamwise and vertical dimensions of the sections are respectively
$d_{x+}=900$ and $d_{y+}= 590$ wall units, although some tests were also carried out for
a streamwise extent of 450 units.
Figure~\ref{longtopic} presents selected motifs for the longitudinal planes 
for $\ntopic=96$.
 As in the cross-flow plane, the dimensions of 
the motifs increase with the wall distance, which is confirmed by
 Figure~\ref{LxLylong}. 
The characteristic dimensions seem 
essentially independent of the  total number of motifs (see also 
next section).
There  is a wide disparity in streamwise characteristic 
dimensions near the wall. 
The motif aspect ratio is highest near the wall and decreases sharply 
in the region $0 < y_+ < 50$.
The vertical dimension increases linearly with the wall distance 
in the region $y_+ > 100$, as well as the streamwise dimension,  with an aspect
ratio of $L_x/L_y$  on the order of 2.

\begin{figure}
\includegraphics[height=5cm]{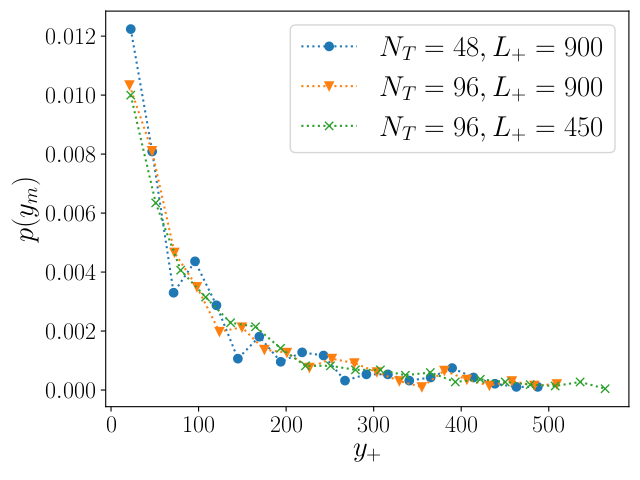} 
\includegraphics[height=5cm]{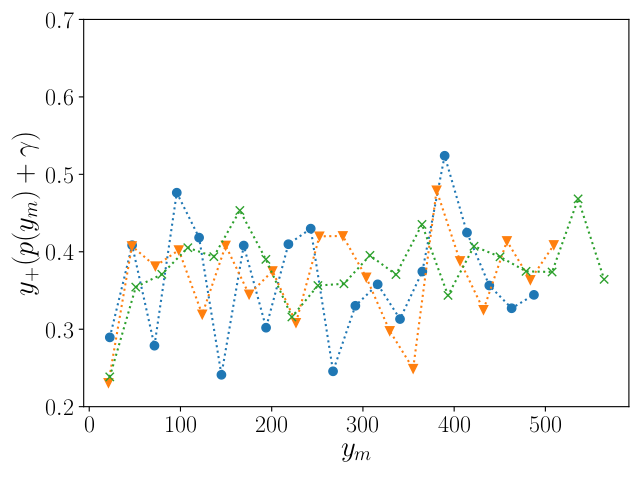} 
\caption{Left: Histogram of the motif location $y^c$; Right: Compensated plot of the histogram for different 
sets of motifs and different subdomains.
The legend is the same for the two figures.}
\label{histzlong}
\end{figure}

Figure~\ref{histzlong} shows the distribution of the motif 
maximum probability location for two different sets  $\ntopic=48, 96,$ and for two domain lengths. 
The shape of the distribution does not appear to change, and again fits well with
the distribution $\proba \simeq \frac{c}{y} - \gamma$ with $c=0.4 $ and $\gamma=-0.0006$
(Figure~\ref{histzlong} right).


\subsection{Sensitivity of the results} \label{Sec_sensitivity}

In this section we examine if and how the characteristics of the motifs
depend on the various parameters of LDA.
We point out that the probabilistic framework of the model makes exact 
comparison difficult, since there is no convergence in the $L_2$ sense,
and the Kullback-Leibler divergence, 
which measures the difference between two distributions
is not a true metric tensor (see Appendix).

The criteria we chose to assess the robustness of the results 
were the characteristic size of the topics 
and the distribution of their locations.
We first examine the influence of various LDA parameters 
on the results obtained for cross-flow sections 
for a constant number of topics $\ntopic=48$.
The reference case corresponded to an amplitude $A=40$,  prior values
of $\alpha=\beta=1/\ntopic$ and a total number of snapshots $\nsnap=800$. 

\begin{figure}
\begin{tabular}{ll}
\includegraphics[height=5cm]{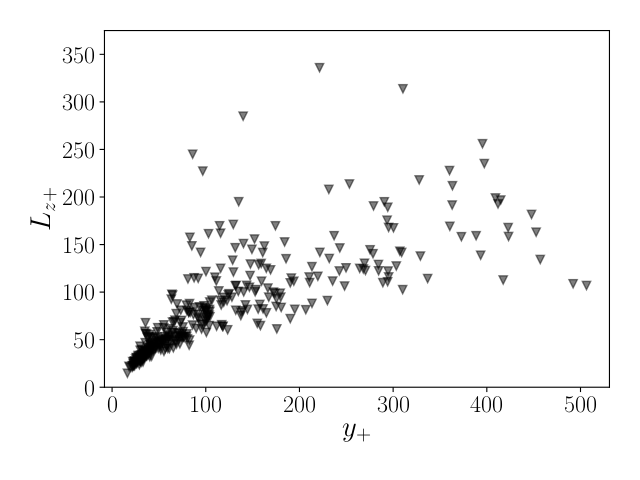} & 
\includegraphics[height=5cm]{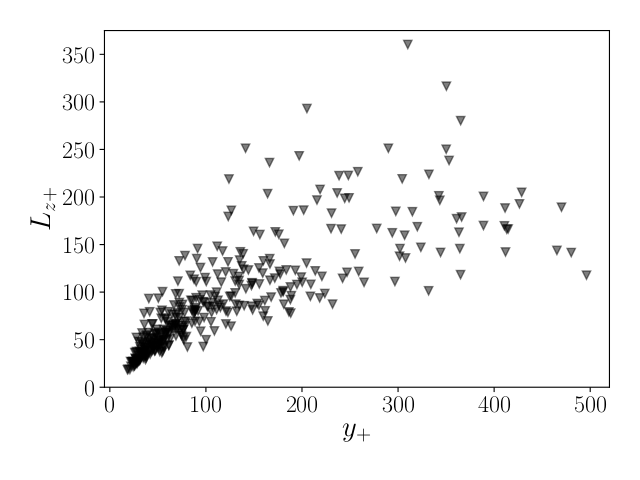} \\ 
\includegraphics[height=5cm]{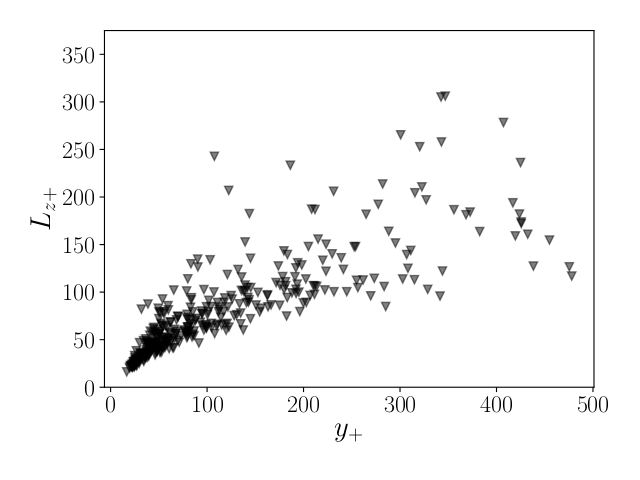} & 
\includegraphics[height=5cm]{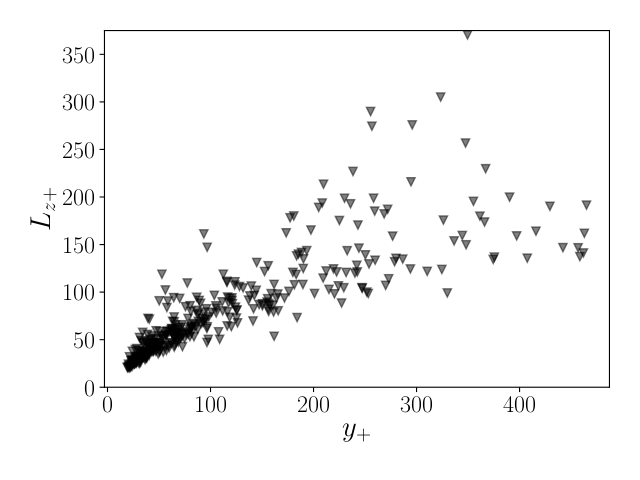} \\ 
\end{tabular}
\caption{ Motif characteristic vertical dimension for $\ntopic=48$.
Top row: Influence of dataset size; $\nsnap$: $\nsnap$= 800 (left), $\nsnap=400$ (right); 
Bottom row: Effect of rescaling  factor; $A=60$ (left); $A=20$ (right).
} 
\label{inputeffect}
\end{figure}

\begin{figure}
\begin{tabular}{cc}
$\alpha= 0.1/\ntopic$,  $\beta= 1/\ntopic$ & $\alpha= 10/\ntopic$,  $\beta= 1/\ntopic$  \\ 
\includegraphics[height=5cm]{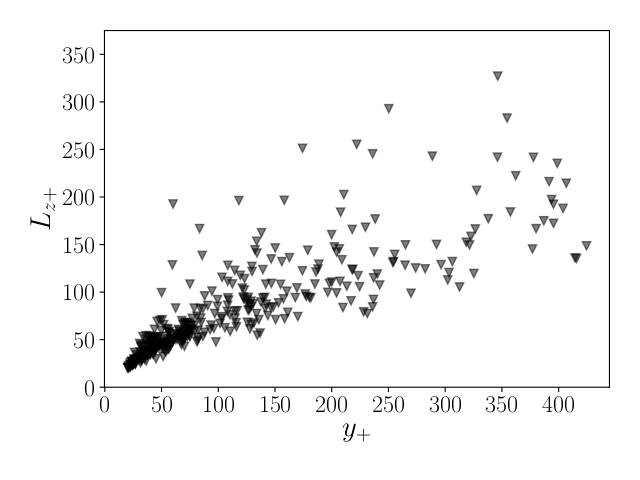} & 
\includegraphics[height=5cm]{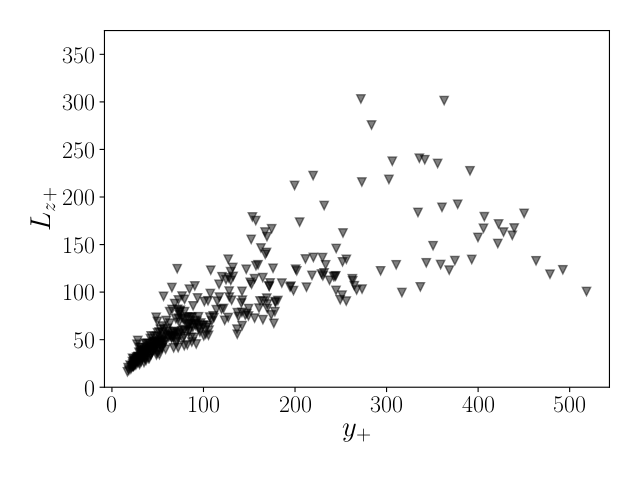} \\ 
$\alpha= 1/\ntopic$,  $\beta= 0.1/\ntopic$ & $\alpha= 1/\ntopic$,  $\beta= 10/\ntopic$  \\ 
\includegraphics[height=5cm]{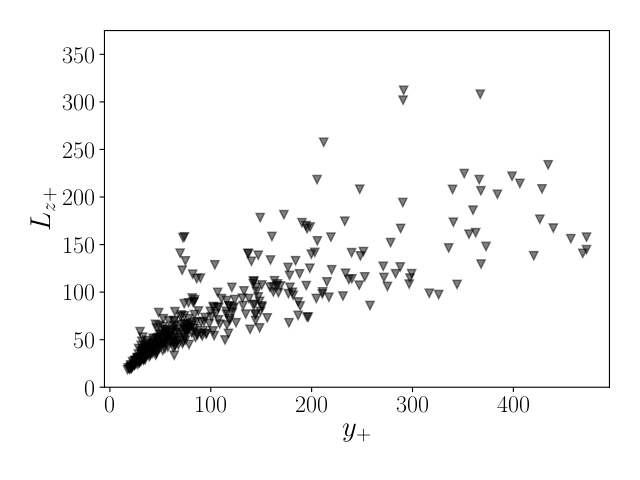} & 
\includegraphics[height=5cm]{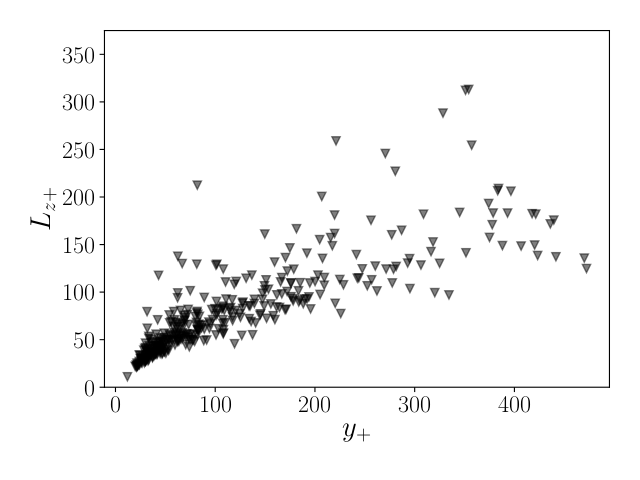} \\ 
\end{tabular}
\caption{
Characteristic vertical motif length for different LDA priors, $\ntopic=48$. }
\label{ldaprior}
\end{figure}

\begin{figure}
\centerline{
\includegraphics[height=7cm]{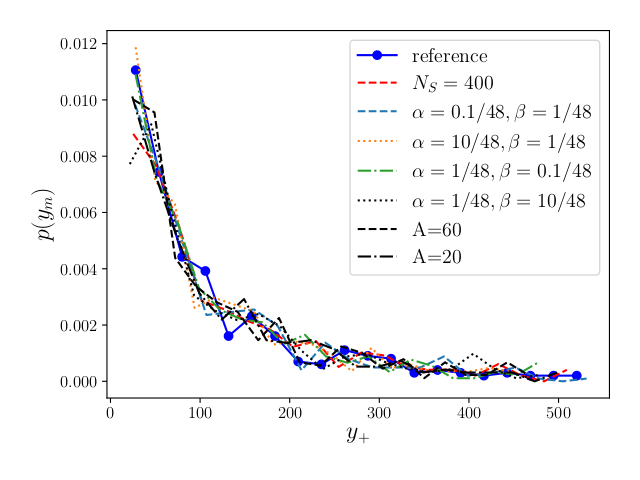}}  
\caption{
Distribution $\proba$ of motif/cell distribution maximum $y_\mathrm{m}$ for different parameters.}
\label{location}
\end{figure}

Figure~\ref{inputeffect} (top row)  shows that the characteristic dimension 
is not modified when the number of snapshots was reduced by 50\%, indicating
that the procedure has converged.
Figure~\ref{inputeffect} (bottom row) shows
 the characteristic vertical dimension $L_y$  of the structures
when the rescaling parameter $A$ was varied. Similar results 
(not shown) were found for $L_z$. 
Although some fluctuations were observed in the individual characteristic 
dimensions,
no significant statistical change was observed. 
Figure~\ref{ldaprior} shows the characteristic dimensions of the structures
for different prior choices for $\alpha$ and $\beta$, which govern the 
sparsity of the representation.
No significant statistical trend was modified when $\alpha$ and $\beta$ were made
to vary within $1/10$ and up to 10 times their default values of $1/\ntopic$. 
Figure~\ref{location} shows that 
the distribution of the maximum location of the motifs 
follows the same inverse law
and does not depend on the choice of parameters chosen for LDA. 

\begin{figure}
\begin{tabular}{ll}
\includegraphics[height=5cm]{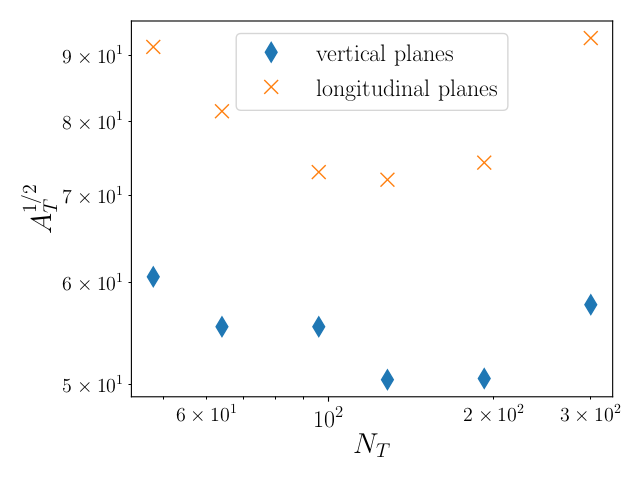} & 
\includegraphics[height=5cm]{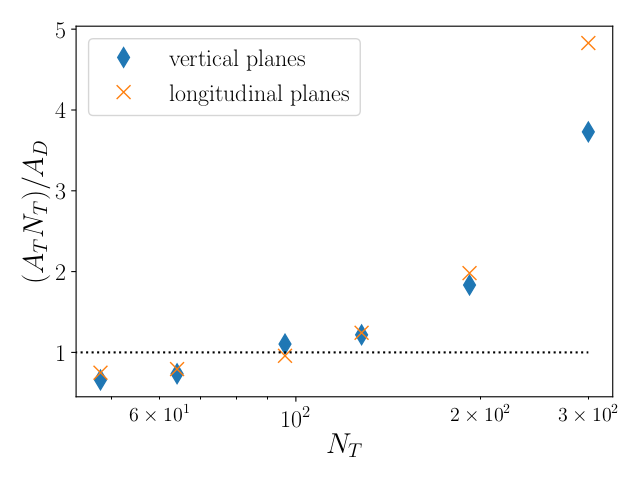} \\  
\end{tabular}
\caption{Left: motif  characteristic dimension $L_T$ for 
different datasets as a function of the number of motifs;
Right: relative fraction of the area captured by the sum
of the topics $\ntopic A_T/A_D$.} 
\label{topicarea}
\end{figure}

We now study the sensitivity of the motifs to the choice of 
$\ntopic$ for both types of vertical planes.
We have seen in the previous sections 
that the motif dimensions appear essentially independent
of the number of motifs considered.
To quantify this more precisely, we 
first define a characteristic motif size $L_T$
as $L_T= \sqrt{\left<A_T\right>}$
where $A_T$ is the area corresponding to the ellipsoid
with the same characteristic dimensions as the motif
and $\left<\cdot\right>$ represents the average over the motifs. 
Figure~\ref{topicarea} summarizes 
how the motif size evolves with the number of motifs
for  both vertical and longitudinal planes.
In all cases, it was found that the characteristic size varies 
slowly around a 
minimal value (Figure~\ref{topicarea}, left), 
and that the  characteristic area of the motif was 
minimum when the sum of the motif 
characteristic areas $\ntopic A_T$ was comparable 
with the total domain area $A_D$ (Figure~\ref{topicarea}, right).   

\subsection{3-D Analysis}

\begin{figure}
\begin{tabular}{ccc}
$x_+=28$ & $x_+=142$ & $x_+=255$  \\ 
\includegraphics[trim=0cm 1cm 8cm 1.1cm, clip, height=6cm]{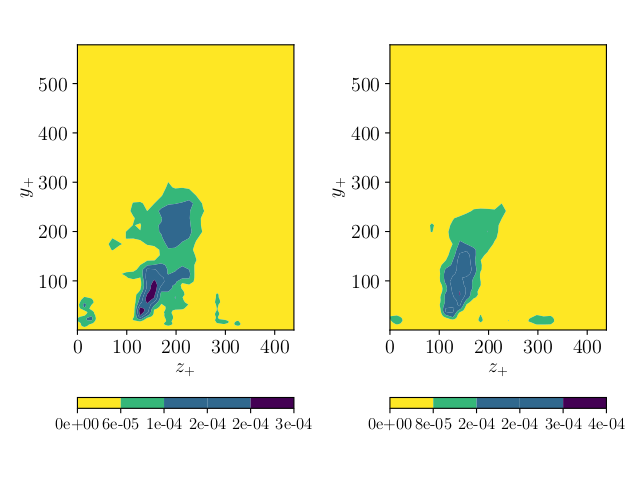}  &
\includegraphics[trim=0cm 1cm 8cm 1.1cm, clip, height=6cm]{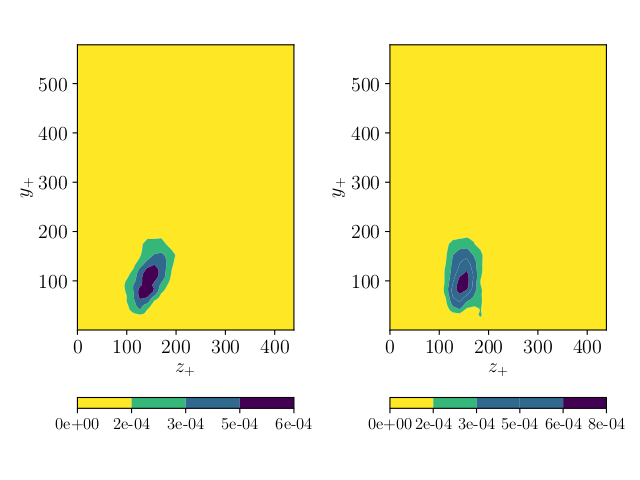}  &
\includegraphics[trim=0cm 1cm 8cm 1.1cm, clip, height=6cm]{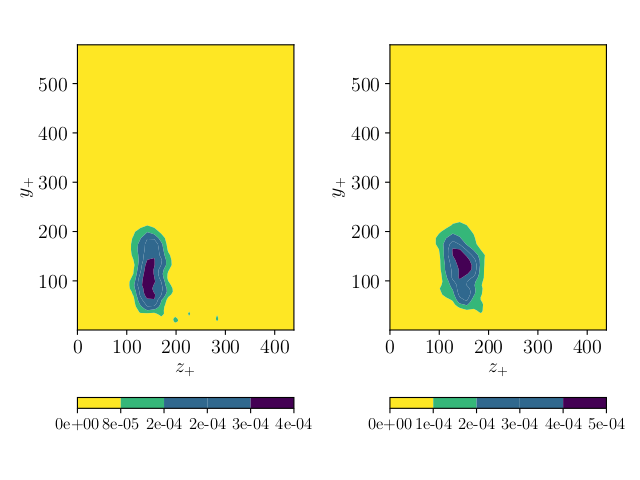}  \\ 
\includegraphics[trim=0cm 1cm 8cm 1.1cm, clip, height=6cm]{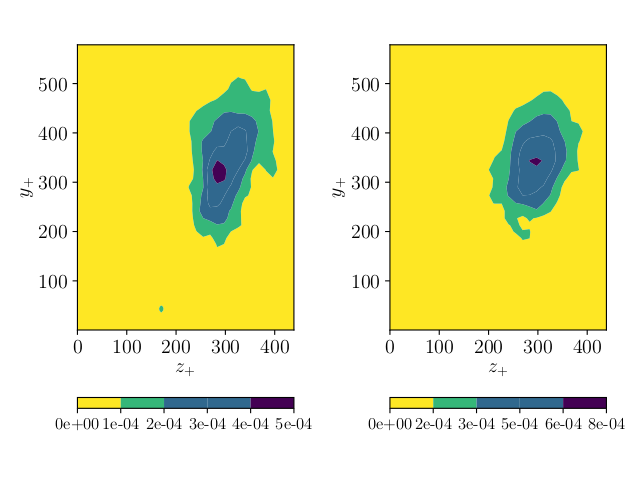}  &
\includegraphics[trim=0cm 1cm 8cm 1.1cm, clip, height=6cm]{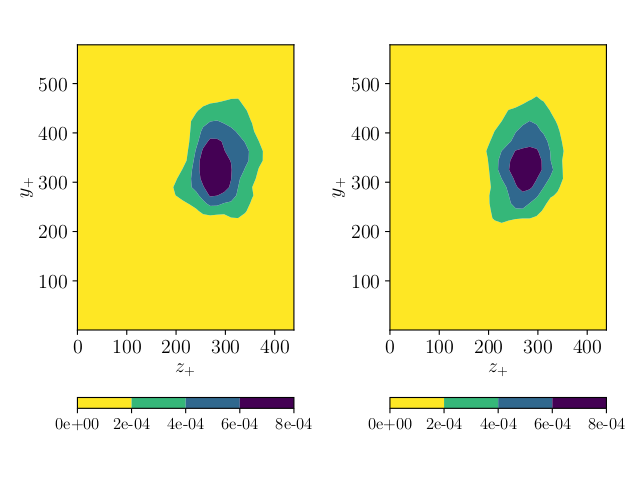}  &
\includegraphics[trim=0cm 1cm 8cm 1.1cm, clip, height=6cm]{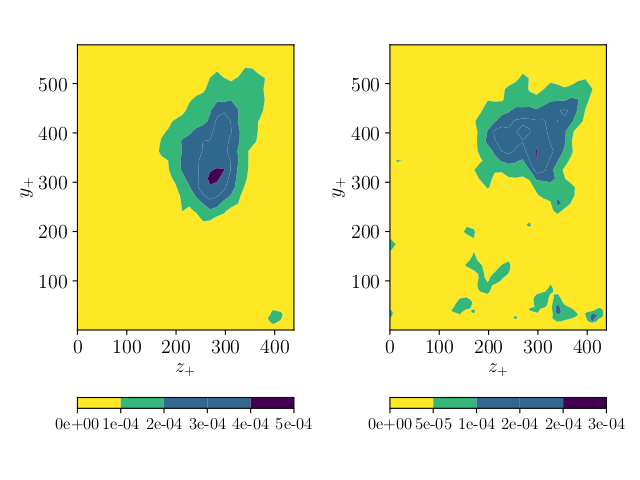}  \\ 
\includegraphics[trim=0cm 1cm 8cm 1.1cm, clip, height=6cm]{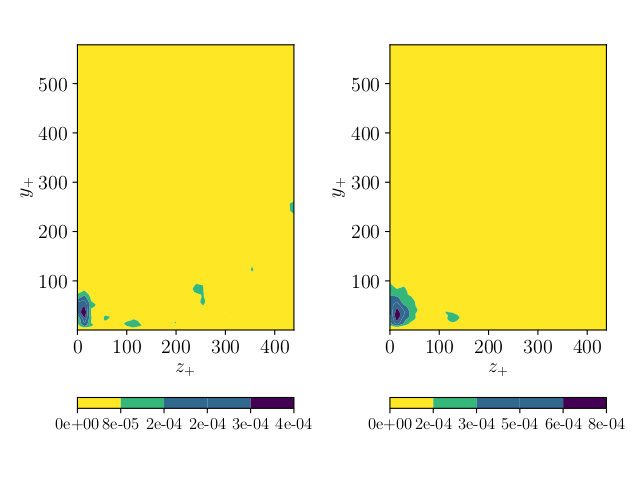}  &
\includegraphics[trim=0cm 1cm 8cm 1.1cm, clip, height=6cm]{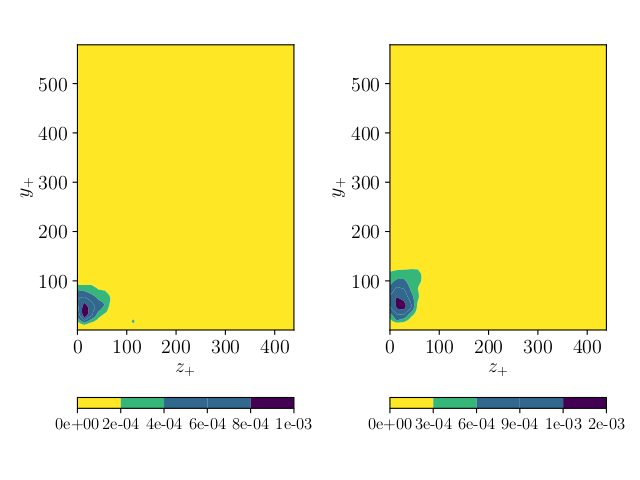}  &
\includegraphics[trim=0cm 1cm 8cm 1.1cm, clip, height=6cm]{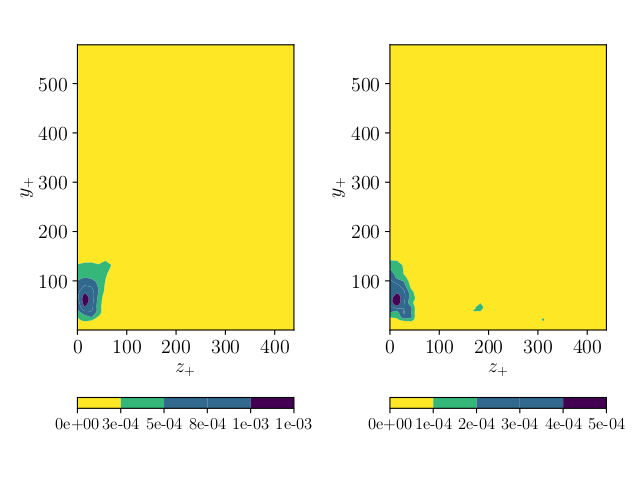}  \\ 

\end{tabular}

\caption{ Cross-sections at different streamwise locations 
of three different 3D motifs obtained for
 $\ntopic=144$; Top row: Motif index $n=34$; Middle row: Motif index $n=7$; Bottom row: Motif index $n=24.$}
\label{topic3d}
\end{figure}

LDA was then applied to a volumic section of the flow of size
$450 \times 590 \times 450$ wall units.
Figure~\ref{topic3d} shows the  cross-sections views of three 3-D motifs.
One can note the streamwise coherence of the topics over different
heights. 
We note that the small dimensions of the volume  may make it difficult to
capture full-length structures, even at this comparatively low Reynolds 
number, and results should be confirmed by a more extensive 
investigation which is outside the scope of this paper.

\begin{figure}
\begin{tabular}{ll}
\includegraphics[height=5cm]{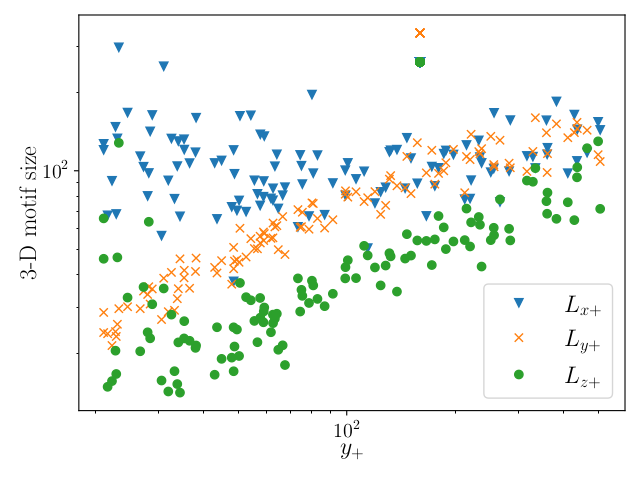} & 
\includegraphics[height=5cm]{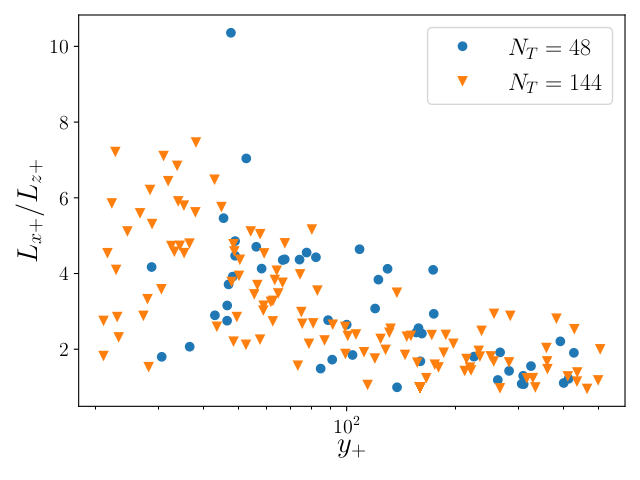} \\ 
\end{tabular}
\caption{ Left: Characteristic dimensions of the 3D motifs, 
$\ntopic=144$;
Right : Evolution of ratio  $L_x/L_z$ with height for $N_T=144$
and $N_T=48$. } 
\label{motif3d}
\end{figure}

The characteristic dimensions of the motifs are reported in Figure~\ref{motif3d}. 
Two different regions can be identified.
For $y_+ < 100$
the region is characterized by a wide distribution  of $L_x$, 
with large values that can extend over the whole domain. 
Some relatively large values of $L_z$ can occasionally be observed. 
For $y_+ < 100$ values of $L_x$ are lower and $L_z$ grows linearly.
$L_y$ appears to grow linearly over both regions.

The ratio between the horizontal dimensions $L_x$ and $L_z$ is reported in Figure~\ref{motif3d} (right).
We can see that the streamwise to spanwise aspect ratio decreases 
over $0 < y_+ < 100$ from an average value of 5 at the wall, which corresponds to the typical aspect ratio of the streaks \citep{kn:dennis15}.
It then decreases more  slowly towards an aspect ratio of about 2 in the region $100 < y_+ < 500.$
This ratio is consistent with results from analysis of POD eigenfunctions in 
\citet{kn:jfe10}, as well as from vortex cluster analysis 
from \citet{kn:delalamo06}.  
3-D motif characteristic sizes are consistent with those obtained for vertical planes, 
which shows that  information about the 3-D organization of the flow can be obtained from
analysis performed on 2-D sections. This is of particular interest as it suggests that 
the LDA method could be usefully applied to PIV experimental data.

\section{Field reconstruction and generation} \label{Sec_reconstruction}

\subsection{Reconstruction}

We now examine how the flow can be reconstructed using LDA.  
{ In all that follows, without loss of generality, we will focus on one of the cross-flow planes examined in Section~\ref{Sec_results}, 
specifically the cross-section at $x=0$ of dimensions $d_{y+}=590$ and $d_{z+}=450$.   }
As described in the algorithm presented in Section~\ref{Sec_LDA}, both the motif-snapshot and the cell-motif distributions can be sampled
for the total intensity 
$N_\isnap = \sum_\ix \snap_{\ix,\isnap}$ in the $\isnap$-th snapshot.
This total intensity is defined as the rescaled integral value of 
the Reynolds stress (digitized and restricted to $Q_-$ events) over the plane.
Since results were found to be essentially independent of the rescaling, 
we can make the simplifying assumption that $N_\isnap$ is large enough so that
the distribution $\wtdistribscal_\imode$ is well approximated by the samples. 
For a given total intensity $N_\isnap$, a reconstruction of the $\isnap$-th snapshot  can  then be obtained at each grid cell $\vecx_\ix$ from 
\[ \tau^\mathrm{R-LDA}(\vecx, t_\isnap) = \frac{1}{A} \snap_i(\vecx)  = \frac{N_\isnap}{A} \sum_{\imode=1}^{\ntopic} \theta_{\imode, \isnap} \wtdistribscal_\imode(\vecx), \]
where 
\begin{itemize}
\item $\wtdistribscal_\imode(\vecx)$ is the motif-cell  distribution,
\item the snapshot-motif distribution $\theta_{\imode, \isnap}$ represents the likelihood of motif $\btopic_\imode$ in the $\isnap$-th snapshot.
\end{itemize}
 
\begin{figure}
\includegraphics[height=7cm]{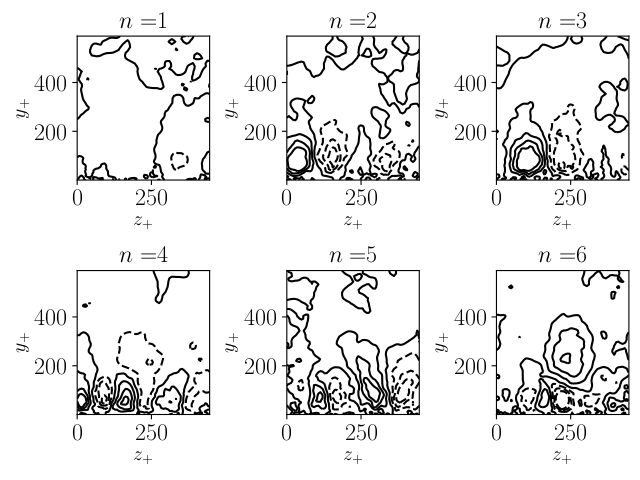}  
\caption{ Contour plot of the first six fluctuating  normalized
POD spatial modes; Contour values go from $-0.03$ to $0.03$.  Negative values are indicated by dashed lines. } 
\label{modepod}
\end{figure}

It seems natural to compare this reconstruction with the POD representation of the field which has a similar expression  
\[ \tau^\mathrm{R-POD}(\vecx,t_\isnap)= \sum_{\imode=0}^{N_\mathrm{POD}-1} \coef_{\imode,\isnap} \phi_\imode(\vecx), \]
where 
\begin{itemize}
\item $\phi_\imode(\vecx)$ are the POD eigenfunctions extracted from the 
autocorrelation tensor $C_{\isnap, \isnap'}$ 
obtained from the $\nsnap$ snapshots,
\item $\coef_{\imode,\isnap}$ corresponds to the amplitude of the $\imode$-th POD mode in the $\isnap$-th snapshot. 
\end{itemize}
The first six fluctuating POD modes are represented in Figure~\ref{modepod}.  
We note that the $0$-th POD mode represents the temporal average of the field.
As expected, the fluctuating POD modes consist of Fourier modes in that spanwise direction (due to homogeneity of the statistics),  and their intensity  reaches a maximum at around $y_+ \simeq 25$. 

If the number of POD modes is equal to the number of motifs $\ntopic$, by construction, POD will provide a better representation of the statistics at least up to second-order \citep{kn:HLB}.
We note that, in terms of computational requirement, POD  
may appear less expensive than LDA, as it requires solving an SVD problem versus implementing an iterative Expectation Maximization 
algorithm \citep{dempster77}.
However the performance
of the EM algorithm can be improved, in particular with online updates
\citep{kn:hofmann99}. 

In terms of storage,
a reconstructed snapshot requires $N_\mathrm{POD}$ modes for POD  and $\ntopic$ topics  
for LDA.
However, storage reduction could be obtained in the case of LDA by filtering 
out the motifs with a low probability $\theta_{\imode, \isnap} $, \ie{}, lower than a threshold $\thres$.
We note that, in this case, it is necessary to store the indices $\imode$ of the motifs as well as the value of $\theta_{\imode, \isnap}$, so that if $n$ modes (resp. topics) 
are kept, storage will consist of $2 n$ variables per snapshot.  
We see that storage reduction can be achieved if the fraction of retained modes $\eta=n/\ntopic$ 
is sufficiently small.
The LDA storage data length per snapshot $2 \eta \ntopic$ should then be compared with the POD data length $N_\mathrm{POD}$.

For $\ntopic=96$, choosing a threshold of $\thres=0.015$ resulted in less than 8\% difference
between the filtered and unfiltered LDA reconstructions (the $L_2$ norm was used). 
 The average value for $\eta$ was $0.2$, which means that 
the number of POD modes that would represent a storage equivalent to that of LDA with
$\ntopic=96$ is $N_\mathrm{POD} \simeq 2 \eta \ntopic \simeq 40$. 
We note that the total storage cost should further take into account the 
size of the LDA basis $\{\btopic_\imode\}_\imode$, which will be larger than the POD basis $\{\bmode_\imode\}_\imode$ since they are respectively equivalent to $\ntopic$ and $N_\mathrm{POD}$ fields.
However efficient storage of the LDA basis can be achieved by making use of the limited 
spatial support of $\btopic_\imode$, in particular for motifs located close to the wall.  

\begin{figure}
\begin{tabular}{cc}
\includegraphics[trim={0 5cm 0 3cm},clip, height=5cm]{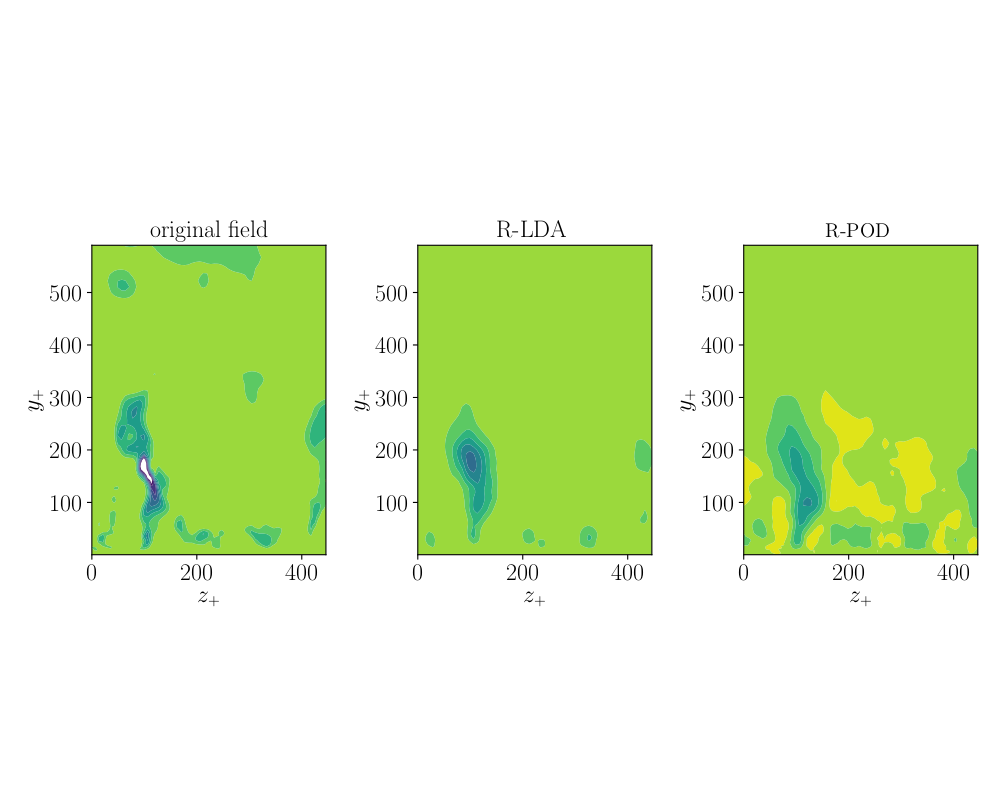} &
\includegraphics[height=4.5cm]{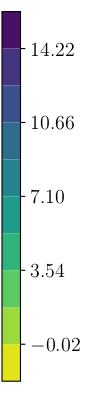} \\
\end{tabular}
\caption{ Instantaneous Reynolds stress field (limited to $Q_-$ events)  
Left: True field; Middle: POD-reconstructed field using 48 POD modes; Right: LDA-reconstructed field using 96 modes. }
\label{comparison}
\end{figure}

\begin{figure}
\centerline{
\includegraphics[height=7cm]{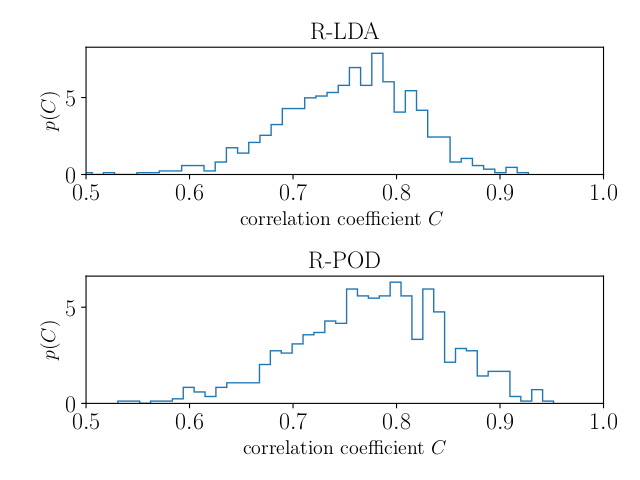}}
\caption{Distribution of the correlation coefficient between each original snapshot and its reconstruction based on LDA (top) or POD (bottom).} 
\label{histcorrelation}
\end{figure}

In the remainder of this section we will compare a filtered LDA reconstruction of 96 motifs (where values of $\theta_{\imode, \isnap}$ lower than
$\kappa=0.015$ are excluded from the reconstruction), with a POD
representation of $N_\mathrm{POD}=48$ modes, which captures about 75\% of the total energy. 
Figure~\ref{comparison} compares an instantaneous field with its LDA reconstruction and its
POD reconstruction.
A more general assessment is provided by Figure~\ref{histcorrelation}, which shows the correlation coefficient
between each snapshot and its reconstruction based on POD as well as that based on LDA.
Although POD appears to be slighty superior, the correlation coefficients 
are very close with respective average values of 0.75 for LDA and 0.77 for POD.  

\subsection{Generation}

LDA is a generative model, so it is straightforward to generate synthetic snapshots by sampling from
distributions $\theta$ and $\wtdistribscal$ for a 
total intensity $N_{\isnap}=\sum_{\ix} f_{\ix, \isnap}$,
which is  modeled as a Poisson process with the 
same mean and standard deviation as the original database. 

\begin{figure}
\centerline{
\includegraphics[height=8.5cm]{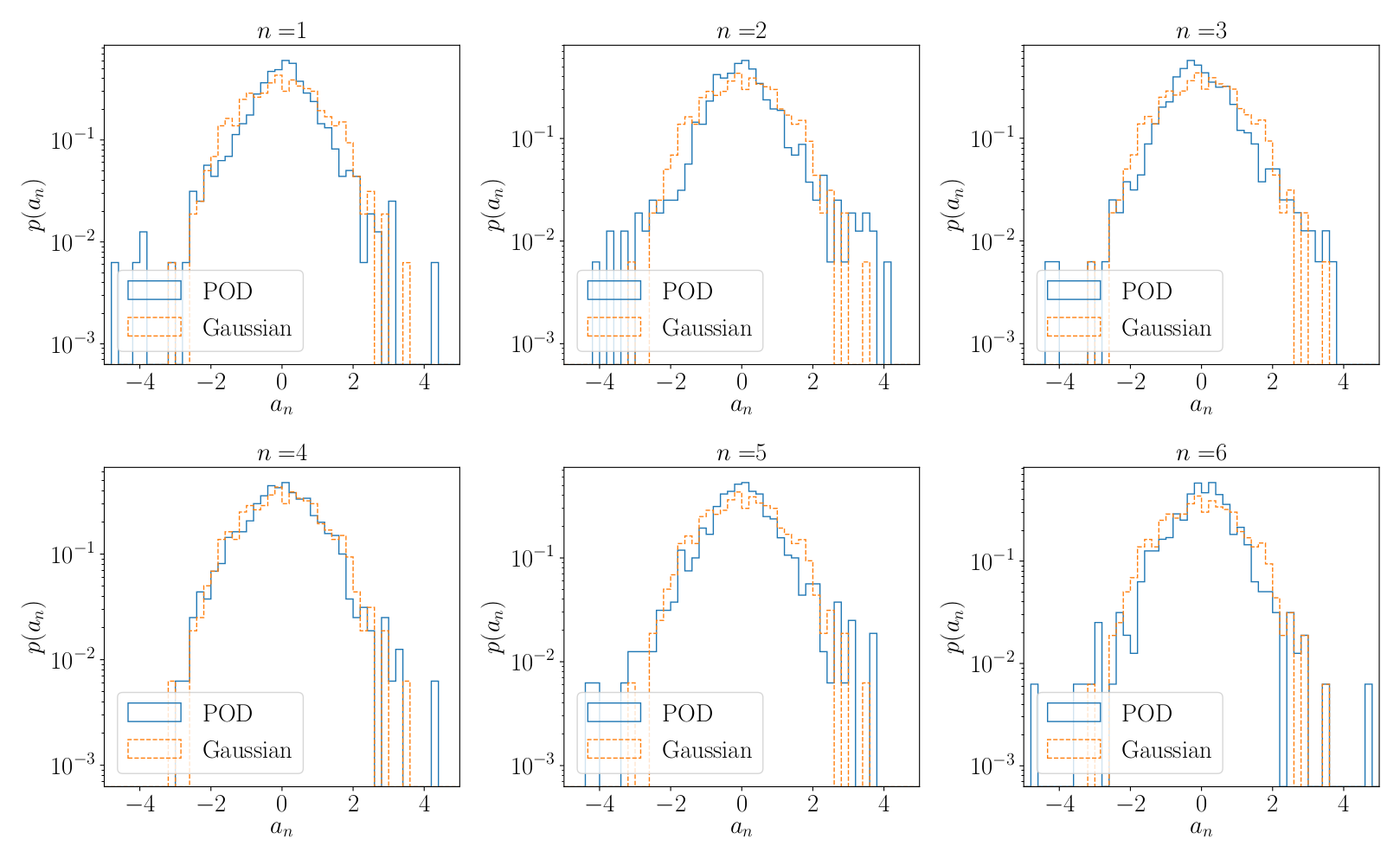} } 
\caption{ Histograms of the normalized amplitudes of the first six 
fluctuating POD modes and comparison with a sampled Gaussian distribution.}
\label{histpod}
\end{figure}

In contrast, POD  is not a  generative model \emph{per se}. 
We will use a simplified version of the  probabilistic extension of POD (PPCA) derived
by  \citet{kn:tipping02}, which is presented in Appendix~\ref{ppca},
where we will make the additional assumption that no noise is present in the model,  
POD-based synthetic fields  will be reconstructed  from deterministic spatial POD modes $\bmode_\imode$ and random  POD amplitudes $\tvecA_{\imode}$ which are assumed to be Gaussian variables.
Examination of Figure~\ref{histpod}, which 
represents the distribution of the first fluctuating POD coefficients 
$\imode \ge 1$, suggests that it is quite acceptable as a first 
approximation to assume  
Gaussian distributions for the amplitudes $\bcoefs_\imode$ --- alternatively, the 
amplitudes could be sampled from the empirical distributions.  
The amplitude of the $0$-th mode, which corresponds to the average of the field over the snapshots, will be assumed to  be constant for all snapshots.

We can therefore compare the databases reconstructed from  and generated  with LDA with those obtained from POD. 
The generated databases consist of $\nsnap$ 
snapshots corresponding to arbitrary instants $\nt_\isnap$.
Overall, the statistics of five different databases can be compared:
\begin{itemize}
\item the true database $\tau_-(y,z,t_\isnap)$ corresponding to the actual values of the $Q_-$ events 

\item the POD-reconstructed (R-POD) or POD-projected database
\[ \tau_-^\mathrm{R-POD}(y,z,t_\isnap) = \sum_{\imode=0}^{N_\mathrm{POD}-1} \coef_{\isnap,\imode} \phi_\imode(y,z), \]
where $\phi_\imode$ are the POD eigenfunctions and $\coef_{\isnap,\imode}$ are the amplitudes of the $\imode$-th POD mode in the $\isnap$-th snapshot.

\item the POD-generated (G-POD) database 
\[ \tau_-^\mathrm{G-POD}(y,z,\nt_\isnap) = \sum_{\imode=0}^{N_\mathrm{POD}-1} \ta_{\isnap,\imode} \phi_\imode(y,z), \] 
where $\ta_{\isnap,0}=\left<\coef_{\isnap,0}\right>$, with $\left<\cdot\right>$ the average over all snapshots and 
$\ta_{\isnap,\imode}$, $\imode \geq 1$, centered Gaussian random variables with variance $\left<\ta_{\isnap,\imode}^2\right>$.

\item the LDA-reconstructed database (R-LDA)
\[ \tau_-^\mathrm{R-LDA}(y,z,t_\isnap) = \frac{N_\isnap}{A} \sum_{\imode=1}^{\ntopic} \theta_{\imode, \isnap} \wtdistribscal_\imode(y,z), \]
where $N_\isnap$ is the   total intensity measured in the $\isnap$-th snapshot,
$\theta_{\imode, \isnap}$ is the distribution of motif $\imode$ on the $\isnap$-th snapshot
and $\wtdistribscal_\imode(y,z)$ is the identified distribution of the cell at $(y,z)$ on 
motif $\imode$. 

\item the LDA-generated database (G-LDA)
\[ \tau_-^\mathrm{G-LDA}(y,z,\nt_\isnap) = \frac{\tN_\isnap}{A} \sum_{\imode=1}^{\ntopic} \ttheta_{\imode,\isnap} \wtdistribscal_\imode(y,z), \]
where $\tN_\isnap$ is the total intensity, which is sampled from a 
Poisson process, 
 $\wtdistribscal_\imode(y,z)$ is the identified distribution of the cell at $(y,z)$ on motif $\imode$ and 
 $\ttheta_{\imode,\isnap}$ is sampled for each $\imode$ from the empirical distribution $\theta_{\imode, \isnap}$ over the snapshots.
\end{itemize}

\begin{figure}
\begin{tabular}{ll}
\includegraphics[height=5cm]{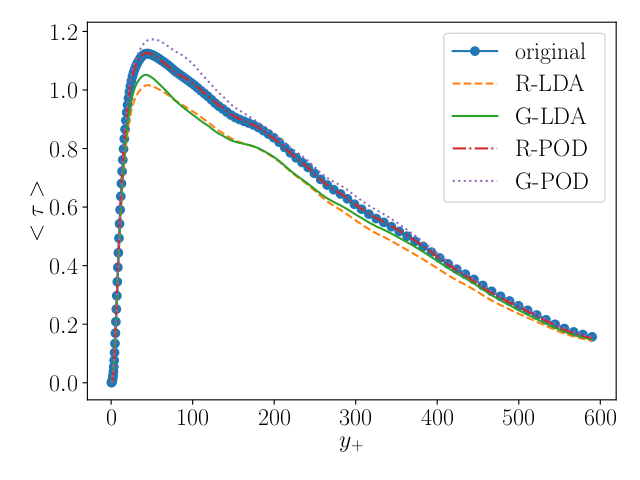} &
\includegraphics[height=5cm]{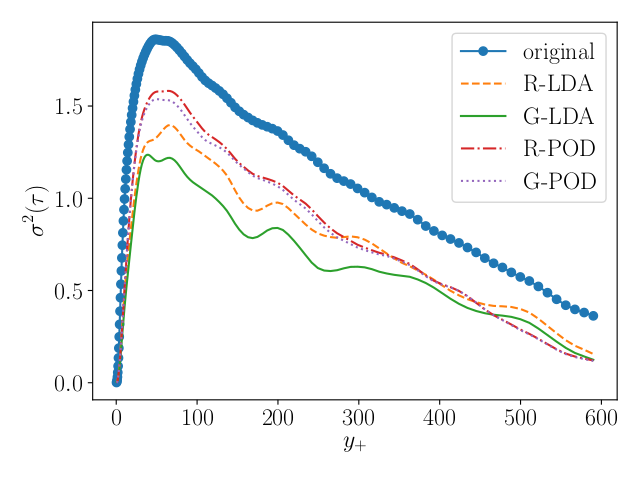} \\
\end{tabular}
\caption{Statistics of the different databases averaged over the spanwise direction and the number of snapshots.
Left: Mean value; Right: Standard deviation.}
\label{database}
\end{figure}

\begin{figure}
\begin{tabular}{ll}
\includegraphics[trim={0 0 0 0},clip, height=5cm,align=c]{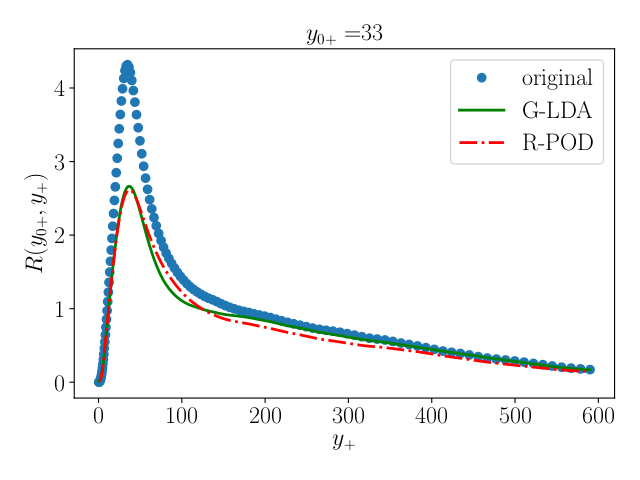} & 
\includegraphics[trim={0 0 0 0},clip, height=5cm,align=c]{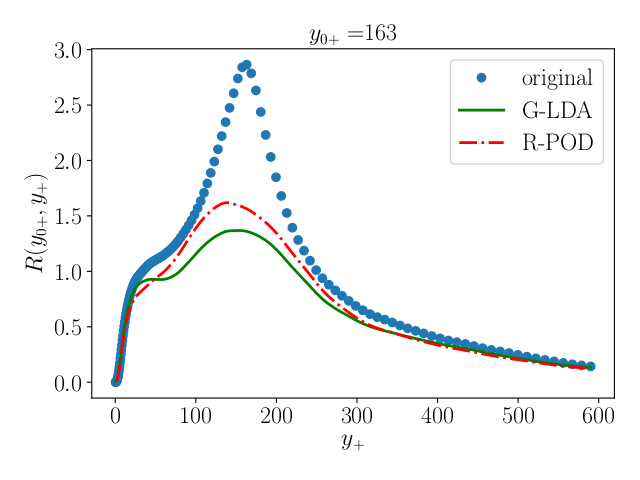} \\ 
\end{tabular}
\caption{Spatial autocorrelation of the Reynolds stress (limited to $Q_-$ events)
in the vertical direction at different heights. The average is taken over snapshots and in the spanwise
direction.}
\label{spatialstructure}
\end{figure}

Figure~\ref{database} shows  the statistics of the different databases as a function of the wall distance.
Averages are taken over all snapshots and in the streamwise direction. 
The mean value of the Reynolds stresses is correctly recovered by all methods.
The second-order statistics are slightly better recovered by the POD-reconstructed
and POD-generated snapshot sets, but both LDA approaches also capture a significant portion of the variance.
The POD databases capture 75\% of the total variance, while the reconstructed and generated LDA databases respectively capture 68\% and 60\% of the variance. 
{Figure~\ref{spatialstructure} shows 
the vertical spatial autocorrelation of $\tau_-$
defined as $R(y, y')=\left<\tau_-(x,y,z,t)\tau_(x,y',z,t)\right>$
(where $\left<\cdot\right>$ represents  an average taken in time and in the spanwise 
position). We can see that the generated LDA autocorrelation 
is very similar to its reconstructed POD 
counterpart, which shows that the LDA synthetic fields capture as much 
as the spatial structure as the POD reconstructed  ones. 
We note that the autocorrelation at large separations is well reproduced
by all datasets.
}

\begin{figure}
\begin{tabular}{lll}
$y_+=19$ &
\includegraphics[trim={0 4cm 0 0},clip, height=3.6cm,align=c]{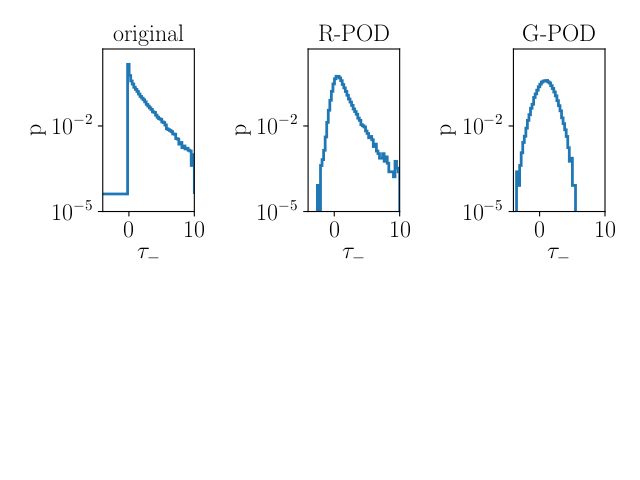} &
\includegraphics[trim={6cm 4cm 0 0},clip, height=3.6cm,align=c]{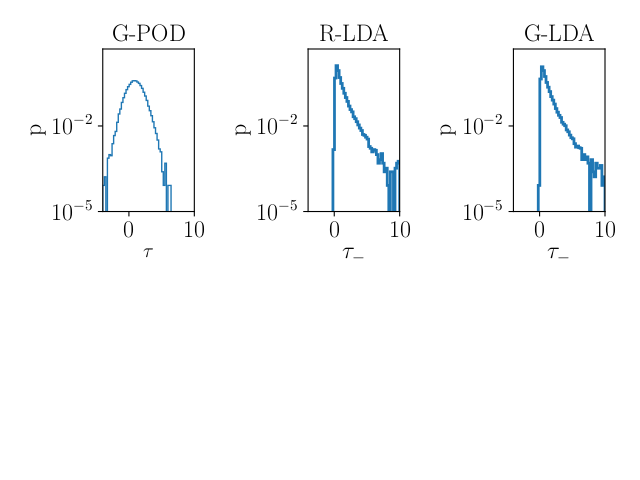} \\
$y_+=61$ &
\includegraphics[trim={0 4cm 0 0},clip, height=3.6cm,align=c]{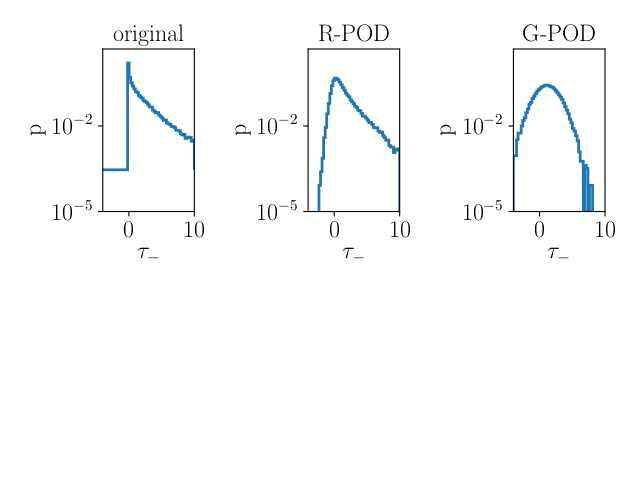} &
\includegraphics[trim={6cm 4cm 0 0},clip, height=3.6cm,align=c]{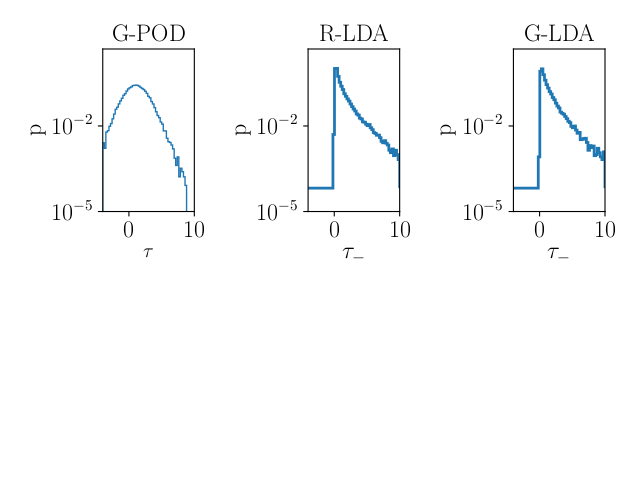} \\
$y_+=157$ &
\includegraphics[trim={0 4cm 0 0},clip, height=3.6cm,align=c]{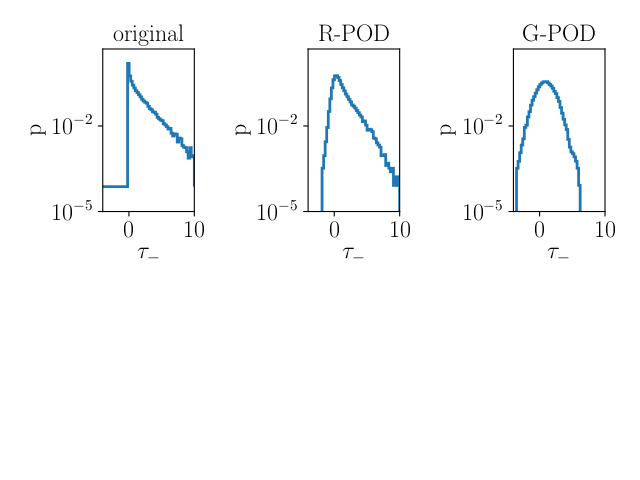} &
\includegraphics[trim={6cm 4cm 0 0},clip, height=3.6cm,align=c]{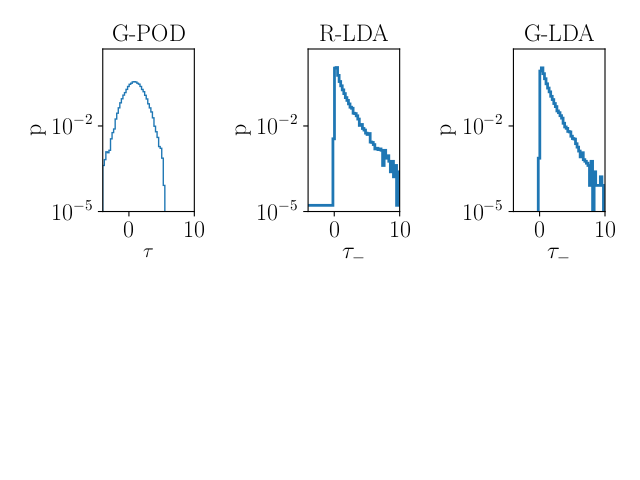} \\
$y_+=343$ &
\includegraphics[trim={0 4cm 0 0},clip, height=3.6cm,align=c]{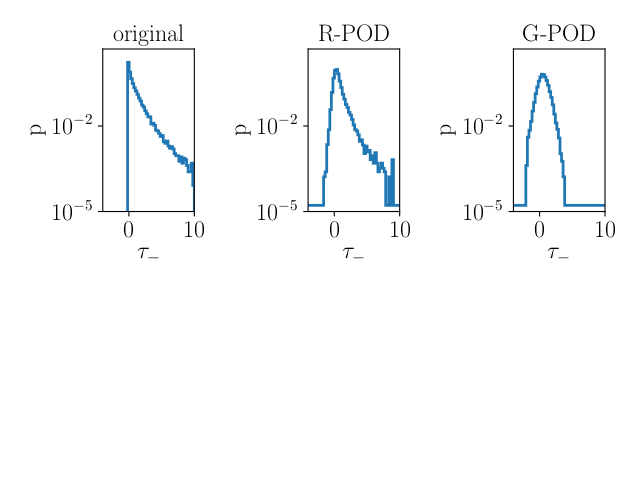} & 
\includegraphics[trim={6cm 4cm 0 0},clip, height=3.6cm,align=c]{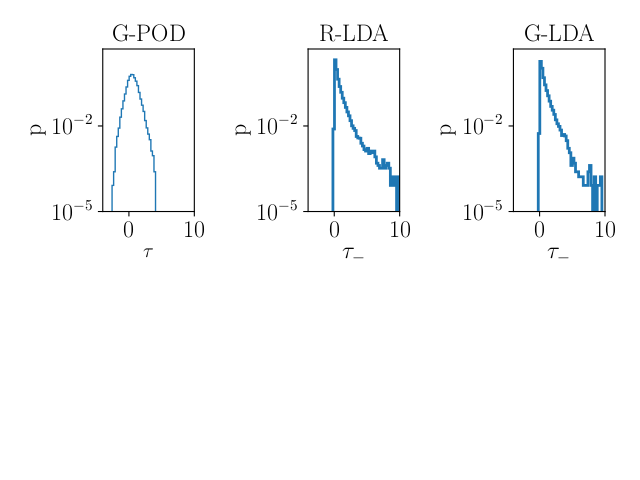} \\
\end{tabular}
\caption{Histograms of the Reynolds stress (limited to $Q_-$ events) 
corresponding to the different databases at different heights.}
\label{histogram}
\end{figure}

Figure~\ref{histogram} shows  histograms of the fields at different heights. 
We note that unlike the LDA approach, which is a non-negative 
decomposition (since it is based on probabilities), 
some negative values are observed for the POD approach, even though the original field values considered 
are always positive. 
We can see that at different wall distances
the POD-reconstructed database reproduces well the distribution of the original database, but
the POD-generated database does not. 
This failure is  due to the fact that although POD amplitudes are uncorrelated by 
construction, they are not independent.
We note that the same failure was observed 
when sampling  the POD coefficients from their data-observed distributions $\coef_{\isnap, \imode}$ 
instead of  Gaussian processes. 
In contrast, both reconstructed and generated LDA methods yield very similar distributions,
which reproduce the main features of the original Reynolds stress values, such as the 
intermittency (sharp peak at zero) and the asymptotic decay for positive values.

\section{Conclusion} \label{Sec_Conclusion}

This paper presents exploratory work about the application 
of Latent Dirichlet Allocation (LDA) to the identification of coherent 
structures in turbulent flows. 
In the probabilistic framework of LDA, 
latent factors or motifs are inferred from a collection of snapshots.
Each snapshot is characterized by a motif distribution, and each motif
itself is distributed over space. 
Implementation was carried out for a scalar field representing  
Reynolds stress $Q_-$ events. 
Evidence of self-similarity was found in the motifs:
the spanwise and vertical dimensions of the motifs increase 
linearly with the wall distance in the logarithmic region, and the number 
of structures evolves inversely with the wall distance. 
This is in agreement with the eddy attached model 
hypotheses.  
The characteristics of the motifs were established to be robust 
with respect to the LDA parameters.   

LDA yields a sparse, efficient
reconstruction of the snapshots that compares reasonably well with POD representation.  Adding in the fact that the motifs have a local spatial support, even when statistics are homogeneous, could  make the LDA representation 
of interest for estimation and control purposes.  
Further, a strong benefit of LDA is its inherent 
generative property, which makes it possible to generate a set of 
synthetic snapshots which is statistically similar to the original one.

The first results obtained with the LDA method open up exciting prospects for data analysis 
and modeling of turbulent flows.
We plan to study larger domains at higher Reynolds numbers in future work.
Moreover, while the investigation was limited to a positive scalar field in the present implementation, 
it would be useful to extend the capabilities of LDA to
fully real, as well as multi-dimensional fields.
Finally, since the technique appears well suited to describe 
intermittent phenomena, it would be 
interesting to apply it to strongly inhomogeneous flow regions such 
as the turbulent/non-turbulent interface 
\citep{kn:philip14} or other types of intermittency \citep{kn:johnson17}.

\section*{Acknowledgments}
This work was supported by the Center of Data Science from the Paris-Saclay University.
Computations were carried out at IDRIS-GENCI (project 02262).
The authors are grateful to the anonymous Referees for their helpful comments on
the first version of the manuscript. 

\section*{Declaration of Interests}
The authors report no conflict of interest.

\appendix

\section{LDA as a factorization method} \label{Sec_MF}

To further shed light on the interpretation of LDA, we now adopt a different viewpoint and briefly explore the connections between the decomposition methods discussed above in the framework of Matrix Factorization (MF). Specifically, we now explain how model decomposition methods, such as POD, K-means and LDA, can be interpreted in terms of Matrix Factorization.

\subsection{Matrix factorization}
Letting $\matsnap\in \R^{\nx\times \ntLM}$ be a data matrix to be approximated, MF consists in the following decomposition:
\begin{equation} 
\matsnap = X Y,
\end{equation}
with $X\in \R^{\nx\times \nmode}$ and $Y\in \R^{\nmode\times \ntLM}$ two real-valued matrices. Compression is achieved whenever $\nmode <\min(\nx,\ntLM)$, which is considered hereafter. MF can be formulated as an  optimization problem:
\begin{equation} 
(X, Y) \in \argmin_{\dumarg{X} \in \admsetX, \dumarg{Y} \in \admsetY} \normLM{\matsnap - \dumarg{X} \dumarg{Y}}{}^2 + \mathcal{R}\left(\dumarg{X}, \dumarg{Y}\right), \label{Eq_FM_canon}
\end{equation}
with $\normLM{\cdot}{}$ a given norm, $\admsetX$ and $\admsetY$ admissibility sets for $X$ and $Y$ respectively, and $\mathcal{R}$ a regularization term.

\subsection{POD-MF equivalence}
Let the singular value decomposition (SVD) of the $\Nx \times \nsnap$ real-valued data matrix $\matsnap$ be
\be
\matsnap = \leftSV \varSigma B^\transpose, \label{SVD}
\ee
with $\leftSV$ and $B$ two orthonormal matrices and $\varSigma$ being diagonal. The Eckart-Young theorem makes precise in which sense this decomposition is optimal, \cite{Eckart1936a}. In particular, it follows that
\be
\leftSV_\nmode, \left(\varSigma B^\transpose\right)_\nmode \in \argmin_{\dumarg{\leftSV}^\transpose \dumarg{\leftSV} = I_\nmode} \normLM{\matsnap - \dumarg{\leftSV} \left(\dumarg{\varSigma B^\transpose}\right)}{F}, \qquad \forall \, \nmode \le \min\left(\Nx, \nsnap\right),
\ee
where $\left(\varSigma B^\transpose\right)_\nmode = \varSigma_\nmode B_\nmode^\transpose$ and with $\leftSV_\nmode$ and $B_\nmode$ the restriction of $\leftSV$ and $B$ to their columns associated with the dominant $\nmode$ singular values $\diag\left(\varSigma_\nmode\right)$.

From Eq.~\eqref{SVD}, it comes
\be
\matsnap \matsnap^\transpose \leftSV = \leftSV_\nmode \varSigma_\nmode B_\nmode^\transpose B_\nmode \varSigma_\nmode^\transpose \leftSV_\nmode^\transpose \leftSV_\nmode = \leftSV_\nmode \varSigma_\nmode^2 = C_\nmode \leftSV_\nmode.
\ee

Refering to Eqs.~\eqref{Eq_eigpb_POD} and \eqref{Eq_empicov}, the diagonal matrix $\varSigma_\nmode^2$ and $\leftSV_\nmode$ then directly identify with the $\nmode$ dominant eigenvalues $\Lambda$ and POD modes $\eigenmat$, respectively. Denoting the Moore-Penrose pseudo-inverse with a $^+$ superscript, the POD projection coefficients are:
\be
A = \eigenmat^+ \matsnap = \eigenmat^\transpose \matsnap = \leftSV_\nmode^\transpose \matsnap = \varSigma_\nmode B_\nmode^\transpose,
\ee
so that the POD decomposition is finally seen to satisfy the following matrix factorization problem:
\be
\eigenmat, A \in \argmin_{\eigenmat^\transpose \eigenmat = I_\nmode} \normLM{\matsnap - \eigenmat A}{F},
\ee
of the form of Eq.~\eqref{Eq_FM_canon} with $\mathcal{R} \equiv 0$ and $\admsetX$ such that $X^\transpose X = I_\nmode$.

\subsection{K-means-MF equivalence}

Clustering is an unsupervised learning technique aiming at identifying groups (clusters) in the data such that data points in the same group have similar features, while data points in different groups have highly dissimilar features.

K-means is one of the simplest and popular clustering methods, \cite{kn:macqueen1967,Lloyd_82}. The algorithm tries to iteratively partition the dataset into $\nmode$ predefined distinct non-overlapping clusters $\left\{ \cluster_\imode \right\}_\imode$. In its standard deterministic version, each data point belongs to only one cluster. The key idea consists in assigning each data point to the closest centroid (arithmetic mean of all the data points that belong to that cluster). The distance is defined in terms of some chosen norm $\normLM{\cdot}{}$.
Setting the number of clusters $\nmode$, the algorithm starts with an initial guess for the $\nmode$ centroids  $\left\{ \centro_\imode \right\}_\imode$, by randomly selecting $\nmode$ data points from the data set without replacement. It then iterates between the data assignment step, assigning each data point $\bsnap_{\isnap}$ to the closest cluster $\cluster_{\imode_\isnap^\star}$ and the centroid update step, which computes the centroid of each cluster:
\begin{align}
\imode^\star_\isnap & \gets \argmax_{1 \le \dumarg{\imode} \le \nmode} \normLM{\centro_{\dumarg{\imode}} - \bsnap_{\isnap}}{}^2, & \forall \: 1 \le \isnap \le \nsnap, \label{Eq_Kmean1} \\
\centro_\imode & \gets \frac{1}{\card{\cluster_\imode}} \sum_{\bsnap_{\isnap} \in \cluster_\imode} \bsnap_{\isnap}, & \forall \: 1 \le \imode \le \nmode. \label{Eq_Kmean2}
\end{align}

K-means is guaranteed to converge to a local optimum but not necessarily to a global optimum. Therefore, we choose to run the algorithm with different initializations of centroids and retain the solution that yielded the lowest loss $\mathscr{L}$:
\begin{align}
\mathscr{L} = \sum_{\imode=1}^\nmode {\sum_{\bsnap_\isnap \in \cluster_\imode}{\normLM{\bsnap_\isnap - \centro_\imode}{}^2}}. \label{Kmeans_loss}
\end{align}

Solving a clustering problem in the $L^2$-sense means finding a set of $\left\{\cluster_\imode\right\}_{\imode=1}^\nmode$ disjoint clusters ($\cluster_\imode \bigcap \cluster_{\imode^{'}} =\{\emptyset\}$, $\imode \ne \imode^{'}$), that minimizes the following cost function: 
\begin{equation} 
\mathscr{L} = \sum_{\imode=1}^\nmode {\sum_{\bsnap_\isnap \in \cluster_\imode}{\normLM{\bsnap_\isnap - \centro_\imode}{2}^2}} = \sum_{\isnap=1}^\nsnap {\normLM{\bsnap_\isnap}{2}^2} - \sum_{\imode=1}^\nmode {\sum_{\bsnap_\isnap, \bsnap_{\isnap'} \in \cluster_\imode}{ n_\imode^{-1} \bsnap_\isnap^\transpose \bsnap_{\isnap'} }}, \label{Eq_clust}
\end{equation}
where $\left\{\centro_\imode\right\}_{\imode=1}^\nmode$ are the cluster centroids, $\centro_\imode := \sum_{\bsnap_\isnap \in \cluster_\imode} {\bsnap_\isnap} \slash n_\imode$, $n_\imode := \card{\cluster_\imode}$.

Let $Y\in\left[0,1\right]^{\ntLM \times \nmode}$ be the normalized cluster indicator matrix, $\by_\imode = n_\imode^{-1 \slash 2} \mathbbm{1}_{\left\{\bsnap_\isnap \in \cluster_\imode\right\}}$.
Disjointedness of clusters implies that columns of $Y$ are orthonormal, $Y^\transpose Y= I_\nmode$.
The clustering problem \eqref{Eq_clust} may now reformulate in terms of $Y \ge 0$ as, \cite{Ding05onthe}:
\begin{align}
Y & \in \argmin_{\dumarg{Y} \ge 0, \dumarg{Y}^\transpose \dumarg{Y}= I_\nmode} \trace{\matsnap^\transpose \matsnap} - \trace{\dumarg{Y}^\transpose \matsnap^\transpose \matsnap \dumarg{Y}}, \nonumber \\
& \in \argmin_{\dumarg{Y} \ge 0, \dumarg{Y}^\transpose \dumarg{Y} = I_\nmode} \normLM{\matsnap^\transpose \matsnap}{F}^2 - 2 \trace{\dumarg{Y}^\transpose \matsnap^\transpose \matsnap \dumarg{Y}} + \normLM{\dumarg{Y}^\transpose \dumarg{Y}}{F}^2, \nonumber \\
& \in \argmin_{\dumarg{Y} \ge 0, \dumarg{Y}^\transpose \dumarg{Y} = I_\nmode} \normLM{\matsnap^\transpose \matsnap - \dumarg{Y} \dumarg{Y}^\transpose}{F}^2.
\end{align}

The Euclidean hard-clustering K-means problem hence stems from an orthogonal non-negative matrix factorization form and the clusters are given by $\centro_\imode = n_\imode^{-1 \slash 2} \matsnap \by_\imode$, $\forall \imode$.

\subsection{LDA-MF equivalence}
We now focus on LDA and discuss the fact that, similarly to POD and K-means, it can also be interpreted as a matrix factorization technique, under certain conditions.

Let us consider the variational LDA flavor, where infering the LDA parameters from maximizing the posterior distribution $p$ is substituted with an approximated posterior $q$, easier to sample from. The inference problem then consists in minimizing the approximation error, which is equivalent to maximizing the Evidence Lower Bound (ELBO) $\loss$:
\be
\loss = \expe_{\mu_q}\left[p\right] - \expe_{\mu_q}\left[q\right].
\ee

Provided suitable approximations in the inference problem are made, and under a symmetric Dirichlet priors hypothesis $\left(\balpha = \alpha \boldone\right)$, \cite{Faleiros_Lopes_2016} have derived an upper bound for the ELBO associated with variational LDA:
\begin{align}
\max \loss \lessapprox \min \sum_{\ix}^\nx { \sum_\isnap^\nsnap{\left(\matsnap_{\ix, \isnap}\log {\frac{\matsnap_{\ix, \isnap}}{{\left(X Y\right)}_{\ix, \isnap}}} + \sum_\imode^\nmode{\mathcal{R}(Y_{\imode,\isnap}, \alpha_\imode)}\right) }}, \label{LDA_NMF}
\end{align}
where $X \geq 0$ and $Y \geq 0$ are variational parameters to infer, normalized as $\sum_{\ix}X_{\ix, \imode}=\sum_{\imode}Y_{\imode, \isnap}=1$, and regarded as normalized probability distributions. 
$\vecx_\imode$ is related to $\bbbeta$ while $\by_\isnap$ is related to the distribution $\btheta_\isnap$ of a document $\bsnap_\isnap$. The term $\mathcal{R}(Y_{\imode, \isnap},{\alpha}_\imode) := (Y_{\imode, \isnap} - {\alpha}_\imode)(\log Y_{\imode, \isnap} - Y_{\imode, \isnap}(\log Y_{\imode, \isnap}-1))$ corresponds to the prior influence and induces sparsity over the document-topic distribution. 

From Eq.~\eqref{LDA_NMF}, it follows that maximizing the ELBO $\loss$ under certain approximations takes the form of a non-negative matrix factorization problem (NMF) of $\matsnap \approx X Y$ expressed in terms of the Kullback-Leibler divergence $D_{I}(\matsnap\|X Y) := \sum_{\ix, \isnap}{ \left(\matsnap_{\ix, \isnap}\log {\frac{\matsnap_{\ix, \isnap}}{{\left(X Y\right)}_{\ix, \isnap}}}-\matsnap_{\ix, \isnap}+{\left(X Y\right)}_{\ix, \isnap} \right)}$, supplemented with a regularization term.
%
%

Details of the derivation are beyond the scope of this paper and one should refer to \cite{Faleiros_Lopes_2016} for a more complete discussion.

\section{Probabilistic PCA/ POD} \label{ppca}

In this section, we 
give a brief review of  Probabilistic PCA (PPCA) \citep{kn:tipping02}
which provides a 
density estimation framework for POD (or PCA/LSA),
under hypotheses that are different from those given in section ~\ref{Sec_PLSA} for PLSA.

We will assume that the data is zero-centered without loss of generality. 
The basic idea  of PPCA is to assume a Gaussian probability  model for the observed data $\snapflu$. 
In that formulation
(see section ~\ref{Sec_PLSA}), the motif-cell matrix $\tilde{\Phi}$
of dimension $N_x \times \nmode$  does not have a probabilistic 
interpretation, but 
relates each noisy observation to a set of $\nmode$ independent 
normalized Gaussian variables  following
\begin{equation}
 \snapflu  = \tilde{\Phi} \tilde{a} + \epsilon      
\label{ppcamodel}
\end{equation}
where the variables $\tilde{a}$ 
are defined to be independent and Gaussian with unit variance
and $\epsilon$ represents noise.  

An important assumption to proceed is that the model for the noise should be isotropic
\[ <\epsilon \epsilon  > = \sigma^2 I, \]
so that all the dependences between the observations are going to be contained 
in $\tilde{\Phi}$. 
On can then show using equation (~\ref{ppcamodel}) that
\[ p(\snapflu) = {\cal N}(0, C ) \]
where 
$C= \tilde{\Phi}^T \tilde{\Phi}  + {\sigma}^2 I $ 
is the observation covariance matrix of dimension $N_x^2$. 

The issue is to determine $\tilde{\Phi}$ and $\tilde{\sigma}$, 
given the observations of $\snapflu$. 
Under the assumption of isotropic  Gaussian noise, 
\citet{kn:tipping02} showed that the maximum likelihood estimators
$\hat{\Phi}$ and $\hat{\sigma}^2$ can be obtained  from 
standard POD analysis on the $\nsnap$ snapshots. 
They showed  that
\begin{equation} 
\hat{\Phi}= \Phi ( \Lambda_{nmode} - \sigma^2  I_{\nmode})^{1/2} R 
\end{equation} 
where $\Phi $ contains the first $\nmode$ eigenvectors of the sampled covariance matrix $\cordual$ 
where $\cordual$ was defined in equation ~\ref{defcordual} (note that
the dimension of $\cordual$ is $\nsnap^2$),
$\Lambda_{\nmode}$ is a diagonal matrix containing 
the $\nmode$ first eigenvalues of $\cordual$ 
and $R$ is an arbitrary rotation matrix.

An estimate  for the error variance can then be given by
\begin{equation}
\hat{\sigma}^2 = \frac{1}{\nsnap - \nmode }\sum_{j=\nmode + 1}^{\nsnap} \eigval_j, 
\end{equation}
which represents the variance lost in the project and averaged over the lost dimensions.

\bibliographystyle{jfm}
\bibliography{all,POD_LDApaper, Biblio}

\end{document}